\documentclass[preprints,article,accept,pdftex,oneauthor]{Definitions/mdpi}

\usepackage{siunitx}
\usepackage{amsmath,amssymb}
\usepackage{booktabs}
\usepackage{array}
\usepackage{float}
\usepackage{bm}
\usepackage{tikz}
\usetikzlibrary{arrows.meta,positioning,shapes.geometric,fit,backgrounds,calc}

\usepackage{longtable} 
\usepackage{array} 
\newcolumntype{N}[1]{%
	>{\raggedright\arraybackslash}p{#1}%
}

\newcommand{\Au}{\ensuremath{^{198}\mathrm{Au}}}
\newcommand{\Ar}{\ensuremath{^{39}\mathrm{Ar}}}
\newcommand{\Twelve}{\ensuremath{T_{1/2}}}

\firstpage{1}

\pubvolume{1}
\issuenum{1}
\articlenumber{0}
\pubyear{2026}
\copyrightyear{2026}

\externaleditor{Firstname Lastname}
\datereceived{9 July 2026}
\daterevised{31 July 2026}
\dateaccepted{4 August 2026}
\datepublished{ }

\Title{Profile Likelihood and Baseline Sensitivity Diagnostics for Digitized Radiation Sensor Decay Datasets}

\Author{{Victor} V. Golovko \orcidA{}}

\AuthorNames{Victor V. Golovko}

\address{%
	Canadian Nuclear Laboratories, Chalk River, {ON K0J 1J0}, Canada; victor.golovko@cnl.ca; Tel.: +1-613-584-3311
}

\abstract{
	Accurate interpretation of radiation sensor decay data is important for environmental monitoring, site remediation, radiation metrology, detector quality assurance, and nuclear data evaluation. When the original gamma spectrometry records are unavailable, a published decay plot may be the only source that can be reanalyzed independently. This study presents a reproducible reduced-data workflow for testing half-life estimates from a digitized \(^{198}\mathrm{Au}\) decay dataset. A weighted exponential fit to the digitized data points reproduces the published room-temperature half-life, indicating that the main decay scale is retained in the figure-level dataset. The analysis then tests how the fitted result changes under plausible figure-level effects, including baseline-like offsets, time-axis reconstruction, finite-window leverage, and ratio-based robustness checks using pairwise summaries and Steiner's most frequent value statistics. The no-offset fit is locally well constrained, but small constant offsets can shift the fitted half-life because the normalization, decay constant, and residual baseline are partly degenerate over the limited time window. Toy Monte Carlo diagnostics show that some estimator shifts are expected for finite-window exponential data. This study does not revise recommended nuclear data or replace the original experiment. Instead, it shows how published radiation sensor decay data can be tested for reproducibility, identifiability, and sensitivity to analysis choices when only reduced or figure-level information is available.
}

\begin{document}
		
	\section{
Introduction
}
	\label{sec:introduction}
	
Accurate interpretation of radiation sensor data is important for environmental monitoring, site remediation, radiation metrology, detector quality assurance, medicine, industry, and nuclear data evaluation. Gamma spectrometry systems are widely used for environmental radioactivity measurements, radioactive material detection, and uranium legacy site assessment~\cite{Fearn2022PhotonPeaks,Guembou2016HPGeSoil,Kunze2022UAVGamma,Carminati2022GammaSpectrometer}.
In many radiation measurement applications, time-dependent count rates, activities, or peak areas are interpreted using exponential decay models to estimate decay constants, half-lives, or related radionuclide time scales~\cite{Gyurky2019Activation,Huizing2020Dosimetry,Shuryak2019RadionuclideKinetics}.
These measurements are sensor-derived quantities, and the fitted half-life can depend on how the detector signal is processed, including photopeak identification, peak integration, background or baseline subtraction, detector calibration, uncertainty assignment, and the available time window~\cite{Bruggeman2018PeakAreas,Tomarchio2023BaselineMDA}.
Standard fits commonly use weighted least squares or maximum likelihood, with the working assumptions that the reported uncertainties are approximately Gaussian and that a single exponential describes the decay data over the analyzed interval~\cite{Brandt2014}.
	
Real measurements are usually less ideal than the simple exponential model assumes. Measured decay data can contain small background or baseline effects that are difficult to separate from the decay signal, especially when the count rate is low near the end of the measurement. The fitted rate can also be affected by the details of peak integration, background subtraction, region-of-interest selection, detector efficiency, dead-time correction, or small changes in the experimental conditions during counting. In addition, the analysis choices made after data collection can influence the fitted half-life. For example, removing early points to avoid short-lived impurities or removing late points because of low counting rates can change the fit’s information balance. Although these effects may be small in absolute size, they can become important when the goal is a high-precision half-life estimate.
	
The problem becomes more difficult when the original experimental data are unavailable. For many legacy measurements and literature-only comparisons, analysts may have access only to a plotted decay dataset, without raw spectra, live-time records, or a detailed analysis log. In that situation, the central question is not what could be achieved with full metadata, but what uncertainty, robustness, and identifiability can be strictly supported by plot-level information alone.
	
This limitation is not restricted to legacy measurements. In many modern nuclear, astroparticle, and high-energy physics experiments, practical barriers---not only access policies---make raw data difficult to share in a directly usable form. The datasets can be very large and often require experiment-specific software, detector calibrations, event selections, metadata, and analysis knowledge to interpret correctly~\cite{adhikari2025,LassilaPerini2021CMSOpenData}. For example, the Dark Matter Experiment using Argon Pulse shape discrimination detector, DEAP-3600, reports about \(150~\mathrm{TB}\) of data for its \Ar{} half-life study, while the Compact Muon Solenoid (CMS) Open Data program had released more than \(2~\mathrm{PB}\) of Run~1 data to external users by 2021~\cite{adhikari2025,LassilaPerini2021CMSOpenData}. Even major A Toroidal LHC ApparatuS (ATLAS) and CMS Higgs-boson papers provide reduced or tabulated results through the High-Energy Physics Data repository, HEPData, rather than the full raw event record~\cite{ATLAS2022HiggsMap,CMS2022HiggsPortrait}. 

In such cases, published figures, summary plots, and tabulated products may be the most accessible material for independent methodological checks. Digitizing figures is therefore not a substitute for full experimental reanalysis, but it can provide a practical and transparent way to test reproducibility, parameter identifiability, and sensitivity to analysis assumptions when full raw data access is unrealistic.	
The isotope \Au{} is a useful case study because its half-life is well established in evaluated nuclear data compilations and has been measured {{repeatedly}}~\cite{HuangKang2016A198}. Therefore, the expected scale is not in doubt, which helps separate questions of analysis robustness from questions of nuclear data evaluation. The data points plotted for the published \Au{} measurement~\cite{spillane2007} provide a case study for showing how parameter correlations, weakly identified nuisance terms, and modest baseline distortions can affect regression-based half-life inference, even when the fit appears visually acceptable.
	
The same isotope also has historical relevance in the literature on possible environmental effects on nuclear decay rates. Early claims that metallic hosts and cryogenic temperatures could modify radioactive half-lives attracted attention both in the specialist literature and in contemporary popular science coverage, where the potential implications for radioactive-waste management and the skepticism of other nuclear physicists were emphasized~\cite{Muir2006}. Modern reviews identify \Au{} as one of the benchmark cases because an early low-temperature measurement reported an anomalous change in the \(\beta^-\)-decay half-life~\cite{spillane2007}, whereas later high-precision measurements did not confirm a significant temperature or host-material dependence~\cite{Belli2026,Goodwin2007AuTemp,Goodwin2010Au2O3,Hardy2010}. In particular, Goodwin et al. found the \Au{} half-life to be unchanged within \(0.04\%\) between room temperature and \(19~\mathrm{K}\), and later found no difference at the \(0.04\%\) level between \Au{} in metallic Au and in \(\mathrm{Au_2O_3}\)~\cite{Goodwin2007AuTemp,Goodwin2010Au2O3}. These later measurements are important context for the present work because they separate physical temperature or host-material effects~\cite{Goodwin2012Dissertation} from the figure-level reconstruction sensitivities studied here.
	
All quantitative results depend on the digitized dataset and the working error model adopted for figure-level analysis. The goal is not to revise the recommended nuclear data values or to retest the cryogenic temperature claim directly. Instead, the goal is to demonstrate a reproducible workflow and quantify the robustness of the reconstructed \Au{} room-temperature digitized data inference under realistic analysis-level perturbations. Because the decay data analyzed here were already background-subtracted in the source publication, additive constants do not represent measured physical background rates in the usual experimental sense. They are used as either conservative diagnostic probes or nuisance parameters representing possible residual baseline bias in digitized, figure-level~data.
	
Recent methodological work provides additional tools for robustness analysis. Pairwise ratio methods remove the absolute normalization from a single-exponential decay curve, while Steiner's most frequent value (MFV) statistic can summarize the resulting heavy-tailed lifetime distributions~\cite{sensors2026pairwiseMFV}. Related work has also shown that confidence intervals and even quoted statistical uncertainties can depend on the fitting or resampling strategy used, even when the fitted central value is stable~\cite{Golovko2025HPB}. In this \Au{} study, these methods were used only as robustness diagnostics. The primary result remains the weighted regression and profile likelihood analysis of the digitized decay data.

Therefore, the central methodological question addressed here is narrow but important: How well can the published room-temperature \Au{} regression result be reproduced from the digitized figure-level dataset, and how sensitive is that reconstruction to baseline, timing, weighting, and other analysis-level choices?

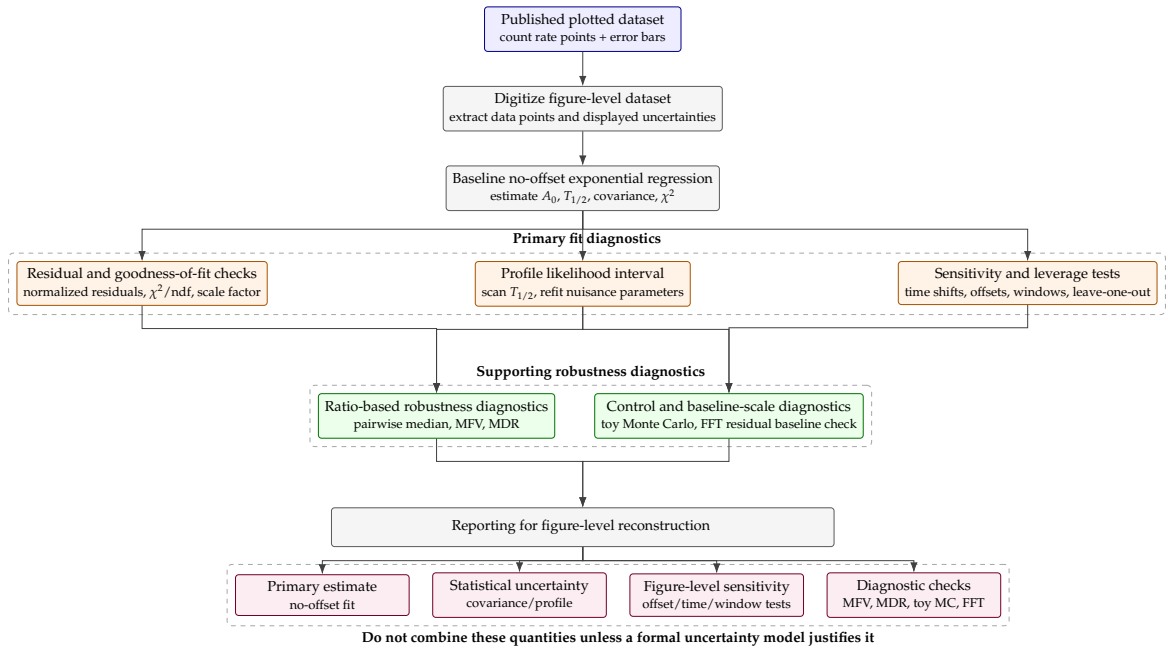
\begin{figure}[H]
	\centering
	\resizebox{0.98\linewidth}{!}{%
		\begin{tikzpicture}[
			node distance=7.5mm and 8mm,
			>={Latex[length=2.2mm,width=1.6mm]},
			block/.style={
				draw,
				rounded corners=2pt,
				align=center,
				inner sep=4pt,
				minimum height=8.5mm,
				font=\small,
				line width=0.45pt
			},
			data/.style={
				block,
				fill=blue!7,
				draw=blue!55!black,
				minimum width=43mm
			},
			process/.style={
				block,
				fill=gray!8,
				draw=black!65,
				minimum width=46mm
			},
			diagnostic/.style={
				block,
				fill=orange!10,
				draw=orange!65!black,
				minimum width=42mm
			},
			support/.style={
				block,
				fill=green!8,
				draw=green!45!black,
				minimum width=42mm
			},
			report/.style={
				block,
				fill=purple!7,
				draw=purple!60!black,
				minimum width=38mm
			},
			arrow/.style={
				->,
				line width=0.55pt,
				draw=black!75
			},
			group/.style={
				draw=black!35,
				rounded corners=3pt,
				dashed,
				inner sep=5pt
			}
			]
			
			\node[data] (published) {
				Published plotted dataset\\
				\footnotesize count rate points + error bars
			};
			
			\node[process, below=of published] (digitize) {
				Digitize figure-level dataset\\
				\footnotesize extract data points and displayed uncertainties
			};
			
			\node[process, below=of digitize] (baseline) {
				Baseline no-offset exponential regression\\
				\footnotesize estimate \(A_0\), \(T_{1/2}\), covariance, \(\chi^2\)
			};
			
			\node[diagnostic, below left=11mm and 39mm of baseline] (residuals) {
				Residual and goodness-of-fit checks\\
				\footnotesize normalized residuals, \(\chi^2/\mathrm{ndf}\), scale factor
			};
			
			\node[diagnostic, below=11mm of baseline] (profile) {
				Profile likelihood interval\\
				\footnotesize scan \(T_{1/2}\), refit nuisance parameters
			};
			
			\node[diagnostic, below right=11mm and 39mm of baseline] (sensitivity) {
				Sensitivity and leverage tests\\
				\footnotesize time shifts, offsets, windows, leave-one-out
			};
			
			\node[support, below=19mm of profile, xshift=-32mm] (robust) {
				Ratio-based robustness diagnostics\\
				\footnotesize pairwise median, MFV, MDR
			};
			
			\node[support, below=19mm of profile, xshift=32mm] (controls) {
				Control and baseline-scale diagnostics\\
				\footnotesize toy Monte Carlo, FFT residual baseline check
			};
			
			\node[process, below=16mm of profile, yshift=-28mm, minimum width=110mm] (hierarchytitle) {
				Reporting for figure-level reconstruction
			};
			
			\node[report, below=6mm of hierarchytitle, xshift=-57mm] (primary) {
				Primary estimate\\
				\footnotesize no-offset fit
			};
			
			\node[report, right=5mm of primary] (stat) {
				Statistical uncertainty\\
				\footnotesize covariance/profile
			};
			
			\node[report, right=5mm of stat] (fig) {
				Figure-level sensitivity\\
				\footnotesize offset/time/window tests
			};
			
			\node[report, right=5mm of fig] (diag) {
				Diagnostic checks\\
				\footnotesize MFV, MDR, toy MC, FFT
			};
			
			\draw[arrow] (published) -- (digitize);
			\draw[arrow] (digitize) -- (baseline);
			
			\draw[arrow] (baseline.south) -- ++(0,-4mm) -| (residuals.north);
			\draw[arrow] (baseline) -- (profile);
			\draw[arrow] (baseline.south) -- ++(0,-4mm) -| (sensitivity.north);
			
			\draw[arrow] (residuals.south) -- ++(0,-5mm) -| (robust.north);
			\draw[arrow] (profile.south) -- ++(0,-5mm) -| (robust.north);
			\draw[arrow] (profile.south) -- ++(0,-5mm) -| (controls.north);
			\draw[arrow] (sensitivity.south) -- ++(0,-5mm) -| (controls.north);
			
			\draw[arrow] (robust.south) -- ++(0,-5mm) -| (hierarchytitle.north);
			\draw[arrow] (controls.south) -- ++(0,-5mm) -| (hierarchytitle.north);
			
			\draw[arrow] (hierarchytitle.south) -- ++(0,-3mm) -| (primary.north);
			\draw[arrow] (hierarchytitle.south) -- ++(0,-3mm) -| (stat.north);
			\draw[arrow] (hierarchytitle.south) -- ++(0,-3mm) -| (fig.north);
			\draw[arrow] (hierarchytitle.south) -- ++(0,-3mm) -| (diag.north);
			
			\begin{pgfonlayer}{background}
				\node[group, fit=(residuals)(profile)(sensitivity),
				label={[font=\small\bfseries]above:Primary fit diagnostics}] {};
				\node[group, fit=(robust)(controls),
				label={[font=\small\bfseries]above:Supporting robustness diagnostics}] {};
				\node[group, fit=(primary)(stat)(fig)(diag),
				label={[font=\small\bfseries]below:Do not combine these quantities unless a formal uncertainty model justifies it}] {};
			\end{pgfonlayer}
			
		\end{tikzpicture}%
	}
	\caption{{Workflow} for reduced-data quality assurance of a digitized radiation sensor decay dataset. The input is the published plotted dataset, meaning the individual count rate points and displayed error bars. The workflow separates the primary no-offset regression estimate from residual checks, profile likelihood intervals, sensitivity tests, and supporting robustness diagnostics. Final reporting separates the primary estimate, statistical uncertainty, figure-level sensitivity scale, and diagnostic~checks.}
	\label{fig:workflow-reduced-data-qa}
\end{figure}

Figure~\ref{fig:workflow-reduced-data-qa} summarizes the reduced-data quality assurance workflow used in this study. The workflow begins with a published plotted dataset and proceeds through regression, residual checks, profile likelihoods, sensitivity scans, and supporting robustness diagnostics. The final step is a reporting hierarchy that keeps primary estimates, statistical uncertainties, figure-level sensitivity scales, and diagnostic checks separate.

{{This} study uses established statistical methods. Its main contribution comes from combining these methods into a reproducible workflow for decay data available only in a published figure. The workflow reports the main regression result, statistical uncertainty, figure-level sensitivity, and supporting checks separately. The reconstruction also provides information that the source article~\cite{spillane2007} did not report. Both fitted parameterizations show strong correlations~\cite{Akoglu2018Correlation} between the normalization and half-life. These strong correlations help explain why baseline shifts, early data points, and the limited observation period can affect the fitted half-life even when the fitted curve looks acceptable. The workflow also shows which conclusions the plotted data can support and which questions require the original spectra, covariance information, and detector records. The workflow does not replace standard experimental data processing. Instead, it provides a structured approach when researchers have access only to published or reduced data and cannot reprocess the original experiment.}

The remainder of this paper is organized as follows. Section~\ref{sec:data} describes the published \(^{198}\mathrm{Au}\) room-temperature decay data used in this study, the digitization procedure, the experimental information available from the source publication, and the limitations of using figure-level data instead of the original spectra and analysis records. Section~\ref{sec:methods} presents the analysis workflow. It begins with the standard weighted exponential fit and residual diagnostics, and then introduces profile likelihood intervals, background and residual offset tests, sensitivity scans, pairwise ratio lifetime estimates, MFV diagnostics, and toy Monte Carlo controls. Section~\ref{sec:results} presents the numerical results. It compares the published half-life, digitized data regression result, profile likelihood intervals, nuisance parameter tests, and pairwise/MFV summaries. It also uses toy data to determine which estimator differences are expected for finite-window exponential data. Section~\ref{sec:discussion} discusses the implications of the results for figure-level half-life reconstruction, including which conclusions are robust and which depend on analysis choices, such as baseline offsets, time-window selection, weighting, and nuisance parameter assumptions. It also outlines future applications to long-duration, high-statistics measurements such as \(^{39}\mathrm{Ar}\). Section~\ref{sec:conclusions} summarizes the main conclusions.
	
	\section{Experimental Data}
	\label{sec:data}
	
The dataset analyzed in this work was obtained by digitizing the individual plotted data points and error bars from the published \Au{} decay data shown in Figure~2 of Ref.~\cite{spillane2007}. Throughout this work, digitization refers to extracting the plotted data points and their uncertainties; the smooth fitted curve and the original raw spectra were not digitized. The figure reports the count rate of the \SI{412}{keV} \(\gamma\) line from \Au{} at \(T=\SI{293}{K}\) as a function of time. This particular measurement has continued relevance because it is included in recent reviews of possible host-material and temperature-dependent decay rate effects, where the original low-temperature \Au{} anomaly and subsequent null measurements are discussed as part of the broader experimental context~\cite{Belli2026}. Digitization was performed using WebPlotDigitizer version 4.8~\cite{rohatgi2024}, a type of graph digitization approach whose reliability has been studied in prior work~\cite{AydinYassikaya2021}. Both axes were calibrated manually using two-point reference selections. Because the source plot used linear axes, the coordinates were extracted using WebPlotDigitizer's native linear-axis mapping.

{The source article~\cite{spillane2007} states that the error bars have a statistical origin, but it does not define their exact meaning. In particular, the article does not confirm that each error bar represents one standard deviation. We therefore use the extracted error bar sizes as relative weights in the fit. We assume that they represent independent uncertainties with an approximately normal distribution, but we cannot confirm this assumption from the published figure.} 

{Digitization can introduce two types of error. First, small differences in point selection can affect individual data points and error bars. The point-picking jitter test examines this effect. Second, the common axis calibration can shift several reconstructed points together and create correlations between them. The published figure does not provide enough information to calculate these correlations. We therefore examine them using time-scale and count rate offset stress tests. We report these tests separately and do not combine them with the displayed error bars.}
	
In the original experiment, the room-temperature decay data were measured over several days using 1-h counting intervals, corresponding to a nominal spacing of
	\[
	\Delta t=\SI{3600}{s}
	\]
between neighboring points~\cite{spillane2007}. This sampling pattern was preserved as closely as possible during digitization. Because the points were extracted from a published figure rather than from the original numerical table, it is not expected that the reconstructed time coordinates will lie exactly on a perfectly uniform 1-h grid. In the digitized dataset, the neighboring time differences remain clustered near \(1~\mathrm{h}\), with reconstructed spacings ranging approximately from \(0.91\) to \(1.14~\mathrm{h}\). These small point-to-point variations are attributed to digitization effects, such as finite graphical resolution, manual point placement, and axis calibration uncertainty, rather than to irregular sampling in the original experiment.
	
The final dataset contains \(n_{\mathrm{pts}}=78\) points. Therefore, the observed variations in neighboring time spacing are treated as reconstruction artifacts. Their possible effect on the fitted half-life is later tested through timing-related perturbations, including small distortions of the reconstructed time scale. These checks verify that plausible timing axis reconstruction errors do not control the inferred half-life.
	
{We cannot determine the exact starting time from the published figure. However, shifting every time value by the same amount does not change the fitted half-life when the fit includes a free normalization parameter. The shift changes only the fitted normalization. The first digitized point occurs at $t_{\min}=116.07~\mathrm{s}$. We shifted the time values by up to $\pm60~\mathrm{s}$ to check the reconstructed coordinates. These shifts did not change the fitted half-life within numerical precision, as expected. In contrast, an error in the time scale changes the fitted half-life. We therefore test time-scale changes separately.}
	
The published Figure~2 curve was fit with \(A(0)\) and \(\Twelve\) as free parameters, where \(A(0)\) is the extrapolated initial count rate and \(\Twelve\) is the half-life. {Spillane et al.}~\cite{spillane2007} reported
	\[
	A(0)=(3.68\pm0.04)~\mathrm{cps},
	\qquad
	\Twelve=(2.669\pm0.017)~\mathrm{d},
	\]
with \(\chi^2_\nu=1.06\) for the individual room-temperature curve~\cite{spillane2007}. The uncertainties shown in the published plot were described as statistically significant.
	
The experimental setup used a high-purity germanium (HPGe) detector with 120\% relative efficiency at \(E_\gamma=\SI{1.3}{MeV}\), positioned on-axis \(\SI{18}{cm}\) from the sample~\cite{spillane2007}. The dead time was monitored with a \(\SI{50}{Hz}\) precision pulser using the method of Anders~\cite{anders1969}, and it was reported to remain below 0.1\%. The continuum under the \(\SI{412}{keV}\) peak was estimated using a local linear background fit. The activated Au foils were \(\SI{0.5}{mm}\) thick, had an area of \(2\times2~\mathrm{cm^2}\), contained less than 1 ppm O and H impurities, and were produced through \(^{197}\mathrm{Au}(n,\gamma)^{198}\mathrm{Au}\) activation. Three room-temperature runs were used by {Spillane et al.}~\cite{spillane2007} to form their reported weighted average. The digitized dataset analyzed here corresponds to one of those individual runs.
	
The final extracted dataset spans
	\[
	t_{\min}=116.07~\mathrm{s}
	\quad\text{to}\quad
	t_{\max}=2.7826\times10^5~\mathrm{s},
	\]
or approximately \(3.2~\mathrm{d}\). The reconstructed time grid and the subsequent script output confirm that the digitized points retain the nominal 1-h spacing of the original measurement.
	
{Digitization introduces known limitations. These include finite graphical resolution, possible correlations among nearby points from shared axis calibration, and possible offsets inherited from the original background subtraction or baseline treatment. Because raw spectra, live-time information, and detailed analysis logs are unavailable for the published figure analyzed here, these effects cannot be reconstructed with a full experimental model. Therefore, the analysis below is framed as a robustness and identifiability study under plausible analysis-level perturbations, not as a definitive re-evaluation of the original~experiment.}

The digitization of published plots is a pragmatic tool for independent consistency checks, sensitivity studies, and methodological investigations when raw data are not accessible. Such analyses do not reproduce the full experimental inference or recover detector-level systematics. Instead, they assess which parts of a reported result are identifiable from reduced, figure-level information and how sensitive those inferences are to reasonable analysis-level assumptions and perturbations.
	
	\section{
Methods
}
	\label{sec:methods}
This section describes the mathematical models and statistical diagnostics used to extract half-lives and quantify uncertainties from a digitized figure-level dataset. The methodology emphasizes analytical techniques that remain viable even when raw spectra and their associated correlation matrices are absent.
	
All analyses treat the digitized count rate measurements as approximately Gaussian, with variances given by the extracted uncertainties \(\sigma_i^2\). Data are treated as statistically independent for fitting purposes. This independence assumption is a practical approximation because digitization can introduce correlations through shared axis calibration, error bar visual extraction, and smoothing in the original plot. Because such correlations cannot be reconstructed without the raw data, we adopt a diagonal error model and later test its adequacy using residual diagnostics, leave-one-out influence tests, and controlled sensitivity studies. Given the moderate-to-high reported count rates and the symmetric error bars in the source figure, the Gaussian approximation provides a reasonable working model for regression and uncertainty diagnostics.

{Normal residuals do not prove that the data points are independent. The common axis calibration and digitization process may create correlations between the reconstructed points. The published figure does not provide enough information to measure these correlations. We therefore keep the diagonal fit as the reference model and use sensitivity tests and residual checks to examine its limitations.}
	
Toy Monte Carlo pseudo-experiments were considered as a validation tool for the reconstruction workflow. In these tests, synthetic decay data are generated from the no-offset best-fit model using the reconstructed time grid and assigned point uncertainties. Such pseudo-data are not treated as an alternative measurement of the \Au{} half-life. Instead, they provide a controlled way to test estimator bias, profile likelihood coverage, and expected response of regression and pairwise/MFV diagnostics under known assumptions.

The analyses below are performed using different levels of inferential approach. The primary figure-level estimate is the weighted no-offset exponential regression, which is checked with a profile likelihood interval. Offset scans and signed offset profiles are used to test sensitivity and practical identifiability. They are not alternative half-life measurements. Pairwise/MFV summaries, MDR smoothing, toy Monte Carlo controls, and the Fourier transform baseline check are supporting diagnostics. They test whether the same half-life scale is recovered under different transformations or idealized controls, but they do not replace the primary regression result.
	\subsection{The Baseline Exponential Decay Model}
	
We begin with the standard exponential decay model because it is the model used in the original fit and provides the reference point for all later diagnostics:
	\begin{equation}
		A(t) = A_0 \exp\left(-\ln 2\,\frac{t}{T_{1/2}}\right),
		\label{eq:baseline}
	\end{equation}
where \(A(t)\) is the model-predicted count rate at time \(t\), \(A_0\) is the extrapolated count rate at \(t=0\), and \(T_{1/2}\) is the half-life. We also use the mean lifetime \(\tau=T_{1/2}/\ln 2\) when presenting some profile likelihoods.
	
The parameters of Equation~\eqref{eq:baseline} are estimated using weighted nonlinear least squares, with weights \(w_i=1/\sigma_i^2\) and \(\sigma_i\) is the uncertainty assigned to the digitized count rate. Fits are performed using the standard Levenberg--Marquardt algorithm~\cite{More1978}. Local covariance and correlation estimates are obtained from the optimum Jacobian using the usual linearized approximation. These covariance estimates are mainly used as diagnostic summaries and as a comparison with profile likelihood intervals, not as the only basis for uncertainty reporting.
	
	\subsection{Goodness-of-Fit and Residual Diagnostics}
	
Goodness-of-fit checks are used to determine whether the fitted curve describes the data within the assigned point uncertainties. For the no-offset exponential fit,
	\begin{equation}
		\chi^2 = \sum_i \left(\frac{A_i-\hat{A}(t_i)}{\sigma_i}\right)^2,
		\qquad
		s^2=\frac{\chi^2}{\mathrm{ndf}}
		=\frac{\chi^2}{n_{\mathrm{pts}}-p},
		\label{eq:chi2}
	\end{equation}
	where \(A_i\) is the digitized count rate at time \(t_i\), \(\hat{A}(t_i)\) is the fitted value, \(\sigma_i\) is the assigned uncertainty, \(n_{\mathrm{pts}}\) is the number of fitted points, \(p\) is the number of fitted parameters, and \(\mathrm{ndf}=n_{\mathrm{pts}}-p\). We also inspect the normalized residuals,
	\[
	r_i=\frac{A_i-\hat{A}(t_i)}{\sigma_i}.
	\]
{These} residuals directly show whether the scatter around the fitted curve is consistent with the quoted error bars.
	
If the assigned uncertainties describe the scatter well, then \(\chi^2/\mathrm{ndf}\) should be close to one. A value that is appreciably greater than one indicates that the data vary more than expected from the individual error bars. This can happen if the point uncertainties are slightly underestimated, if a small unmodeled effect exists, or if the neighboring points are not fully independent. In this case, the best-fit parameters are not changed, but the scale factor should be used to increase the statistical uncertainties
	\[
	s=\sqrt{\chi^2/\mathrm{ndf}}.
	\]
	
{The} standard uncertainty-scaling approach is commonly used in particle and nuclear data evaluations when the observed scatter is larger than expected. The fit should not be reported as more precise than the data justify: if the residuals are wider than the assigned uncertainties predict, the fitted uncertainty is widened accordingly. This scaling is only used to calibrate the reported statistical uncertainty. It does not remove or model any possible underlying systematic effect.
	
	\subsection{Profile Likelihood Interval Construction}
	
The strong correlations between \(A_0\) and \(T_{1/2}\), together with the mild nonlinearity of the exponential model motivate uncertainty estimates beyond the linearized covariance matrix. Therefore, we construct profile likelihood curves for each parameter of interest by fixing that parameter on a grid and refitting the remaining parameter at each point.
	
The resulting profiles are reported as follows:
	\begin{equation}
		\Delta\chi^2 = \chi^2 - \chi^2_{\min},
		\label{eq:dchi2}
	\end{equation}
where \(\chi^2_{\min}\) is the fit’s global minimum. For one parameter of interest, we use the conventional threshold \(\Delta\chi^2=1.0\), which corresponds approximately to a 68.27\% confidence interval, as a practical guide. This threshold cannot be interpreted as an exact coverage guarantee. It is used as an approximate interval marker under the Gaussian error working model and the assumption of regular likelihood behavior.

{We use two uncertainty scales and keep them separate. The profile interval uses the extracted point uncertainties without adjustment and follows the conventional $\Delta\chi^2=1$ threshold. We compare this interval with the unscaled covariance uncertainty. We report the covariance uncertainty after applying the residual scale factor separately.}
	
Baseline offsets are a well-known problem in exponential analysis. In the general model, the constant term can bias the recovered decay constant if it is not removed or explicitly modeled. Istratov and Vyvenko (1999) reviewed several approaches for treating such offsets, including Fourier-based extraction, algebraic baseline estimates, differencing, exponential depression, and regularized methods~\cite{IstratovVyvenko1999}.
	
Several methods can be used to examine the baseline offsets in the exponential decay data. In this study, MDR smoothing and the Fourier transform check are used only as diagnostics and not as alternative primary half-life estimators. MDR is a ratio-based smoothing test and is described in Appendix~\ref{app:mdr}. Because the plotted \Au{} data points had already been background-subtracted, MDR was applied to the net count rate data and used only to check whether the reconstructed half-life scale was stable under a ratio-based transformation. Appendix~\ref{app:fft_baseline} provides a separate Fourier transform check of the residual baseline, including the definition and interpretation of \(B_{\mathrm{FFT}}\), is given in Appendix~\ref{app:fft_baseline}.
	
	\subsection{Physical Background Versus Residual Offset}
	
Additive constant terms have different meanings depending on how the data were processed. If raw count rates are directly fitted, a constant term represents a physical background rate and should be nonnegative. The dataset considered here was digitized from published plotted decay data points that had already undergone net peak extraction and background subtraction. In this case, a constant term is better interpreted as a residual baseline offset from imperfect subtraction or digitization, and it may have either sign. Therefore, we explicitly distinguish between a nonnegative physical background \(B_{\mathrm{bg}}\) and a signed residual offset \(B_{\mathrm{off}}\).

For completeness, a nonnegative physical background model was also examined. This check is described in Appendix~\ref{app:auxiliary_checks} because it is not part of the primary inference for the background-subtracted curve. Its role is to demonstrate how a constrained additive term behaves when it is weakly identified. The more relevant baseline diagnostic for the processed digitized data is the signed residual offset model below, in which either sign is allowed for the offset.

	\subsection{Signed Residual Offset Nuisance for Background-Subtracted Data}
	
For background-subtracted net peak data, the offset model is used instead
	\begin{equation}
		A(t)=A_0\exp\left(-\ln 2\,\frac{t}{T_{1/2}}\right)+B_{\mathrm{off}}, \qquad B_{\mathrm{off}}\in\mathbb{R},
		\label{eq:offmodel}
	\end{equation}
where \(B_{\mathrm{off}}\) represents a residual baseline offset that may arise from several non-exclusive sources. The original peak analysis may not remove the background perfectly, and small systematic shifts can also occur during peak area extraction. When the published figure is digitized, additional small shifts can be introduced, for example, through axis calibration, finite graphical resolution, or a consistent point-picking bias. Because these effects can move the reconstructed count rates either upward or downward, the residual offset is treated as a signed quantity rather than a strictly positive background term.
	
This residual offset is treated as a nuisance parameter because, over a finite observation window, a small constant bias can partially mimic a change in the decay constant and shift the inferred half-life. At fixed \(T_{1/2}\), the model is linear in \((A_0,B_{\mathrm{off}})\), so the nuisance solution can be obtained by weighted least squares without iterative nonlinear refitting. This gives a continuation-stable profile in \(T_{1/2}\) and a direct estimate of the nuisance trajectory \(\hat{B}_{\mathrm{off}}(T_{1/2})\).
	
In addition to the fully unconstrained fit (``LS: free''), a weakly regularized variant (``LS: prior'') in which a Gaussian penalty stabilizes \(B_{\mathrm{off}}\):
	\begin{equation}
		\chi^2_{\mathrm{tot}}(T_{1/2}) = \chi^2_{\mathrm{data}}(T_{1/2}) + \left(\frac{B_{\mathrm{off}}-B_0}{\sigma_B}\right)^2.
		\label{eq:offsetprior}
	\end{equation}
	{Here}, \(B_0\) is the weighted mean residual from the no-offset fit, and \(\sigma_B\) is conservatively chosen from the empirical scatter of \(\hat{B}_{\mathrm{off}}\) near the profile minimum, optionally multiplied by a softening factor. This penalty is only used as a regularization device. It is not a physical prior on the experimental background.
	
	\subsection{Sensitivity and Robustness Studies}
	
Sensitivity studies test whether the inferred half-life is stable under plausible analysis choices that cannot be resolved from the digitized figure alone. These tests include uniform additive offsets applied to the count rates, perturbations to the time axis that represent absolute timing shifts or small-scale distortions, and refits restricted to truncated time windows to probe the influence of early and late data. We also perform leave-one-out refits to assess point leverage and repeat selected fits under uncertainties’ global rescaling.
	
These sensitivity studies are qualitative stress tests rather than a formal systematic uncertainty  budget. Their purpose is to reveal possible instability modes, identify which portions of the data exert the greatest influence on the fit, and determine whether conclusions depend strongly on specific modeling choices or subsets of points.

{We chose conservative limits for these stress tests rather than treating them as measured reconstruction uncertainties. The $\pm0.05~\text{counts per second}$ offset equals about \mbox{1.5 times} the weighted root mean square of the residuals, $0.0330~\text{counts per second}$, and about 1.7 times the median displayed point uncertainty, $0.0286~\text{counts per second}$. We also used the $\pm0.5\%$ change in the time scale only to test sensitivity, not to describe the uncertainty of the original timing system. The fitted result changes smoothly across these ranges, so the endpoints define the limits of the stress tests.}
	
	\subsection{Relationship with Robust Estimators}
The observed regression sensitivities are then related to the known behavior of alternative, heavy-tailed robust estimators. Comparing regression results to these estimators provides valuable context for finite-window biases.
	
The observed sensitivity patterns are interpreted in the context of the known properties of alternative half-life estimators. One important alternative is the pairwise or statistical sampling approach developed by Silverman, in which ratios of decay measurements at two different times are used to eliminate the unknown normalization parameter~\cite{Silverman2014}. This idea was later applied in practice to radioactive half-life measurements by Lorusso et al., who demonstrated that pairwise sampling can provide useful half-life estimates from real decay data~\cite{Lorusso2017}. For a pure exponential decay, such ratios produce a set of pairwise lifetime estimates that directly depend on the decay constant rather than on the fitted amplitude. However, because these estimates involve ratios and logarithms of noisy measurements, their distribution can be heavy-tailed even when the original count rate uncertainties are approximately Gaussian.
	
Robust central value methods, such as Steiner's most frequent value (MFV) estimator, are useful in this setting because they summarize the dense central cluster of a heavy-tailed distribution while down-weighting extreme values~\cite{Steiner1973,Steiner1988}. The MFV can be derived by approximating the unknown data distribution with a Cauchy substituting distribution and minimizing the associated Kullback--Leibler information loss. This leads to coupled equations for the location and scale parameters, or dihesion~\cite{Golovko2025Biomolecules}. Because Steiner's original MFV publications~\cite{Steiner1991,Steiner1997} are not readily accessible to many readers, we cite a recent open-access publication that presents the working MFV equations and demonstrates their use in nuclear data and half-life analyses~\cite{Golovko2025Biomolecules,sensors2026pairwiseMFV}. In the HPGe time-series application, background-subtracted detector data were converted into pairwise lifetime estimates and then summarized with MFV statistics for a \(^{97}\mathrm{Ru}\) half-life analysis. The pairwise distribution had a narrow central peak with long tails, making a robust central value estimator useful~\cite{sensors2026pairwiseMFV}.
	
The MFV method has also been used for other analyses, including groundwater inverse modeling, geophysical well-log clustering, astrophysical abundance and cosmological parameter compilations, neutron lifetime evaluation, and radiation sensor background studies, which shows its broader role as a robust estimator for non-Gaussian or outlier-prone datasets~\cite{Szucs2006MFVGroundwater,Szabo2021MFVWellLogs,Zhang2017MFVLi,Zhang2018MFVHubble,Zhang2022MFVNeutron,Zhang2024MFVM87,ZhangChen2026KLTNetMFV,Golovko2023SensorsMFV}.

As an implementation check, the MFV algorithm was implemented in {R}~(version~4.5.1)~\cite{RCoreTeam2026} and applied to the neutron lifetime dataset analyzed by Zhang et al.~\cite{Zhang2022MFVNeutron}.
The deterministic MFV obtained here was \(881.16~\mathrm{s}\), which is in agreement with the published value. Using a non-parametric bootstrap with 5000 resamples, the 68.27\% confidence interval was \([878.83,883.44]~\mathrm{s}\), which is close to the published interval \([878.81,883.41]~\mathrm{s}\). The 95.45\% interval was \([877.72,885.64]~\mathrm{s}\), which was also consistent with the published interval \([877.72,885.61]~\mathrm{s}\). This confirms that the MFV implementation and non-parametric bootstrap uncertainty procedure reproduce the published neutron lifetime benchmark within the Monte Carlo variation.
	
In the present \Au{} analysis, MFV-based estimation is not used as the primary estimator. The study reproduces and diagnoses the published two-parameter regression fit from digitized figure data and does not replace it with a ratio-based estimator. Instead, the pairwise and MFV approach is used as the methodological context. It motivates the emphasis on residual checks, profile likelihood scans, leave-one-out tests, and baseline offset perturbations when assessing the robustness of the regression result.
	
	\section{
Results
}
	\label{sec:results}
	
This section summarizes what can be recovered from decay data that are available only as a published figure. First, we report the baseline weighted fit using the exponential model in Equation~\eqref{eq:baseline}. Then, we assess whether the digitized uncertainties are consistent with the observed scatter and quantify uncertainty and robustness using profile likelihoods, nuisance parameter profiling, and controlled perturbations.

	\subsection{Baseline Weighted Fit}
	\label{sec:baseline_fit}
	
The weighted nonlinear least-squares fit of the digitized \Au{} decay data to \mbox{Equation~\eqref{eq:baseline}} reproduces the overall trend of the published data. The fit gives
	\begin{equation*}
		T_{1/2}=(230{,}575 \pm 1{,}477)~\mathrm{s}
		= (2.6687 \pm 0.0171)~\mathrm{d},
	\end{equation*}
with an extrapolated initial rate
	\begin{equation*}
		A_0=(3.7333 \pm 0.0103)~\mathrm{cps}.
	\end{equation*}
	
{The} quoted uncertainties are one-standard-error statistical uncertainties estimated from the local covariance matrix of the weighted least-squares fit, where the standard error of each fitted parameter is obtained from the square root of the corresponding diagonal variance term~\cite{Dowd2014}. They describe the statistical uncertainty under the adopted no-offset and diagonal error models. {The reported half-life uncertainty of $0.0171~\mathrm{d}$ includes the scale factor $s=1.201$, which accounts for the extra scatter in the residuals. Without this adjustment, the covariance uncertainty is about $0.0142~\mathrm{d}$. This difference must be considered when comparing the covariance uncertainty with the $\Delta\chi^2=1$ profile interval based on the original extracted uncertainties.} They exclude possible figure-level effects from digitization, residual baseline offsets, background subtraction details, or model choice assumptions.
	
The reconstructed half-life agrees closely with the value reported for the original \(T=293~\mathrm{K}\) Figure~2 fit by Spillane et al.~\cite{spillane2007}. The original publication reported
	\[
	A(0)=(3.68\pm0.04)~\mathrm{cps},
	\qquad
	T_{1/2}=(2.669\pm0.017)~\mathrm{d},
	\]
with \(A(0)\) and \(T_{1/2}\) fitted as the free parameters. This agreement shows that the digitized dataset reproduces the central value of the published room-temperature decay curve fit.
	
The fitted half-life and initial rate should not be treated as fully independent quantities. The fitted parameter values were reported in the original publication, but the covariance matrix and the correlation between \(A(0)\) and \(T_{1/2}\) were not reported. In this figure-level reanalysis, using the same no-offset exponential model, the local covariance matrix gives the following:
	\begin{equation*}
		\rho(A_0,T_{1/2})=-0.831 .
	\end{equation*}
{This} is new diagnostic information from reconstruction.
	
The negative correlation has a simple meaning. Over the finite time range covered by the digitized decay data, a shorter half-life can partly compensate for a larger  initial rate \(A_0\). Similarly, a longer half-life can partly compensate for a  smaller \(A_0\). These nearby combinations of \(A_0\) and \(T_{1/2}\) can give decay curves that are very similar over the measured interval. The fitted half-life should therefore be interpreted together with the fitted normalization and their covariance, rather than as an isolated one-parameter result.

This parameter correlation also explains why the local covariance matrix uncertainty should be checked against a profile likelihood. The covariance matrix provides a useful first estimate of the statistical uncertainty. However, it only describes the local, approximately parabolic behavior of the objective function near the best-fit point. This approximation can be incomplete when parameters are strongly correlated or when a parabola does not represent the objective function well over the relevant range~\cite{Venzon1988}. Profile likelihood methods provide a more direct check by scanning one parameter of interest and re-optimizing the remaining nuisance parameters at each scan point. This construction is widely used for confidence intervals (CI) in the presence of nuisance parameters and has good frequentist coverage in many physics applications~\cite{RolkeLopezConrad2005,Raue2009ProfileLikelihood,Cowan2011,PDGStatistics2023}.
		
\begin{figure}[h]
	\includegraphics[width=0.99\linewidth]{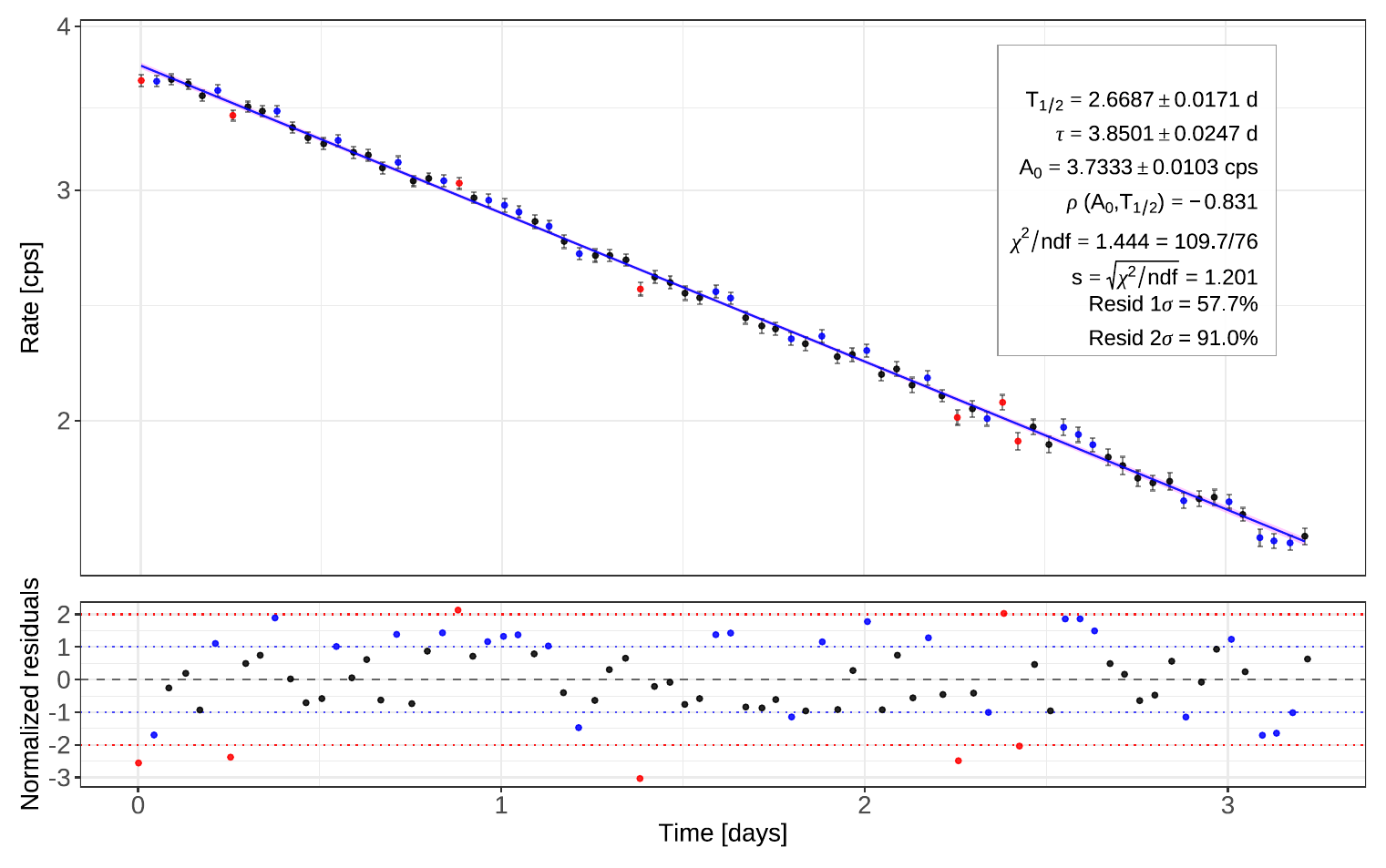}
	\caption{{Baseline} weighted fit of Equation~\eqref{eq:baseline} to the digitized \Au{} data points extracted from the published figure. The (\textbf{{top}}) panel shows the fitted decay curve and digitized count rate data with error bars. The (\textbf{{bottom}}) panel shows the normalized residuals of 		\(r_i=[A_i-\hat{A}(t_i)]/\sigma_i\). The same color coding is used in both panels: black for \(|r_i|\le1\), blue for \(1<|r_i|\le2\), and red for \(|r_i|>2\). This color mapping links each data point in the fit panel to its lower-panel residual band. The annotation summarizes the best-fit half-life, mean lifetime, initial activity parameter, parameter correlation \(\rho(A_0,T_{1/2})\), goodness-of-fit \(\chi^2/\mathrm{ndf}\), and scale factor \(s=\sqrt{\chi^2/\mathrm{ndf}}\).}
	\label{fig:baseline}
\end{figure}

Figure~\ref{fig:baseline} shows that the fitted exponential curve follows the digitized data over the entire time range. The normalized residuals are centered around zero and do not show a time-dependent trend. This supports the use of the two-parameter no-offset exponential model as a reasonable baseline description of the digitized data. The detailed goodness-of-fit and uncertainty-scale checks are discussed in Section~\ref{sec:gof}.

	\subsection{Goodness-of-Fit and Uncertainty Calibration}
	\label{sec:gof}
	
The normalized residuals provide a practical check of whether the assigned point uncertainties have the correct scale. The goodness-of-fit statistic was calculated as the sum of the squared normalized residuals, Equation~\eqref{eq:chi2}. With \(n_{\mathrm{pts}}=78\) fitted points and \(p=2\) fitted parameters, the number of degrees of freedom is
	\[
	\mathrm{ndf}=n_{\mathrm{pts}}-p=78-2=76 .
	\]
{The} fit gives
	\[
	\chi^2/\mathrm{ndf}=109.709/76=1.444,
	\]
corresponding to the scale factor
	\[
	s=\sqrt{\chi^2/\mathrm{ndf}}=1.201 .
	\]
{This} means that the normalized residuals are about \(20\%\) wider than expected if the assigned point uncertainties fully describe the point-to-point scatter.
	
The residual-band counts show the same pattern. Of the 78 points, 45 fall within \(\pm1\sigma\), 26 fall between \(1\sigma\) and \(2\sigma\), and seven fall outside \(\pm2\sigma\). For normally distributed residuals with correctly scaled uncertainties, the expected counts are approximately 53.25, 21.20, and 3.55 points, respectively. Therefore, the observed counts show fewer points in the central band and more points in the outer band than expected. This does not mean that the fit has failed, but it does show moderate extra scatter relative to the stated uncertainties.
	
The residual normality tests do not show strong evidence for a non-normal residual shape. The Shapiro--Wilk, Anderson--Darling, and Lilliefors tests give the following:
	\[
	p_{\mathrm{SW}}=0.1167,\qquad
	p_{\mathrm{AD}}=0.1362,\qquad
	p_{\mathrm{LF}}=0.4242 ,
	\]
	respectively~\cite{ShapiroWilk1965,AndersonDarling1952,Lilliefors1967}. These values are all above 0.05. Therefore, the tests do not reject the normal error assumption. However, this does not mean that the assigned uncertainties have the correct scale.

{The normality tests examine only the shape of the residual distribution. They do not show whether nearby residuals are related. We therefore checked the residuals in time order for repeated signs and patterns between nearby points. The residual plot did not show a clear pattern, but a visual check cannot prove that the residuals are independent. Because the correlations introduced by digitization are unknown, this study treats possible correlation between residuals as a limitation.}
	
If all point uncertainties are rescaled uniformly as \(\sigma'_i=s\sigma_i\), the fitted central values do not change because all weights are multiplied by the same constant factor. The rescaling changes the reported statistical uncertainties and not the best-fit half-life or normalization. The scale factor is useful: it prevents the statistical uncertainty from being reported as more precise than the residual scatter supports.
	
This uncertainty-scaling check tests the overall size of the assigned errors. However, it does not test every possible source of bias. A reduced chi-square value close to one would not rule out small systematic effects such as a residual baseline offset, imperfect background subtraction, or correlations between neighboring digitized points. For this reason, the later profile analyses test whether the inferred half-life remains stable when additional nuisance terms are introduced.
	
The original Figure~2 fit by {Spillane et al.}~\cite{spillane2007}  reported \(\chi^2_\nu=1.06\), with the plotted uncertainties described as statistical only. Since no separate systematic uncertainty budget was provided for that individual fitted curve, the quoted \(\pm0.017~\mathrm{d}\) uncertainty is treated here as the reported statistical fit uncertainty, not as a full total uncertainty.
	
The small difference between the original reported goodness-of-fit and the reconstructed value is not interpreted as a disagreement. This analysis uses digitized points and reconstructed point uncertainties, whereas the original numerical data, exact peak area values, and full weighting procedure are unavailable. In addition, statistical uncertainties from nonlinear fits can depend on the minimization algorithm and covariance estimation procedure. A similar behavior has been observed in half-life reanalyzes where different fitting tools gave nearly identical central values but different statistical uncertainties~\cite{Golovko2025HPB}.
	
The decay data analyzed here is treated as background-subtracted. Therefore, the constant additive terms have different meanings in the following checks. The nonnegative term, \(B_{\mathrm{bg}}\ge0\), is used only as a conservative stress test for a possible remaining physical background. A signed term, \(B_{\mathrm{off}}\), is used later as a more flexible diagnostic for residual baseline shifts that could arise from imperfect subtraction, digitization, or peak-extraction~effects.
		
As an additional check of the fitted parameter coupling, the fit was repeated using an equivalent parameterization of the exponential model. Instead of writing the model in terms of the initial rate \(A_0\),
\[
A(t)=A_0\exp(-t/\tau),
\]
we used
\[
A(t)=\frac{\mathcal{N}_0}{\tau}\exp(-t/\tau).
\]
{Here}, \(\tau=T_{1/2}/\ln2\) is the mean lifetime, and \(\mathcal{N}_0\) is an effective normalization scale. In this count rate fit, \(\mathcal{N}_0=A_0\tau\) in the chosen time units. The reparameterized fit gives \(\mathcal{N}_0=1.2419\times10^6\) when \(\tau\) is expressed in seconds, and it gives the same equivalent initial rate, \(A_0=3.7333~\mathrm{cps}\). This scale should not be interpreted as the number of radioactive atoms or source decays unless the measured count rate is explicitly connected to the source activity through the detector efficiency, photon emission probability, counting geometry, live time, and any relevant correction factors. These factors directly influence the standard gamma spectrometry activity and efficiency relations~\cite{Golovko2022Efficiency,Golovko2024Scintillation}.

Changing from the \((A_0,\tau)\) form to the \((\mathcal{N}_0,\tau)\) form is useful as a diagnostic test, but it does not remove the normalization--lifetime coupling caused by the limited observation window. The reparameterization helps show whether the fitted lifetime is stable under an equivalent model form. This is consistent with Box's view that robust statistical modelling should examine how conclusions change under plausible changes to the working model~\cite{Box1979}.

This reparameterization is similar to the physical parameterization discussed in Ref.~\cite{sensors2026pairwiseMFV}, where changing from an \((A_0,\tau)\) form to an \((\mathcal{N}_0,\tau)\) form changed the correlation structure while leaving the fitted lifetime essentially unchanged. In the present \Au{} analysis, the reparameterized fit also gives the same half-life and equivalent initial rate as the baseline fit, but the correlation between the fitted normalization and the half-life remains strong:
	\[
	\rho(\mathcal{N}_0,T_{1/2})=0.936 .
	\]
	
{Thus}, changing the parameterization does not remove the coupling between the scale parameter and the decay constant. This supports the interpretation that the parameter dependence is a feature of fitting a finite-window exponential decay curve, rather than only an artifact of using \(A_0\) as the amplitude parameter.
	
	\subsection{Profile Likelihood of the No-Offset Model}
	
The no-offset profile likelihood checks whether the local covariance matrix uncertainty is consistent with the full one-parameter scan. To achieve this, we construct profile likelihood curves in \(\Delta\chi^2\). One parameter is fixed at a sequence of trial values, and the remaining parameter is re-optimized at each point. This gives a direct view of how rapidly the weighted least-squares objective worsens as the parameter of interest moves away from its best-fit value.
	
\begin{figure}[h]
		\includegraphics[width=0.99\linewidth]{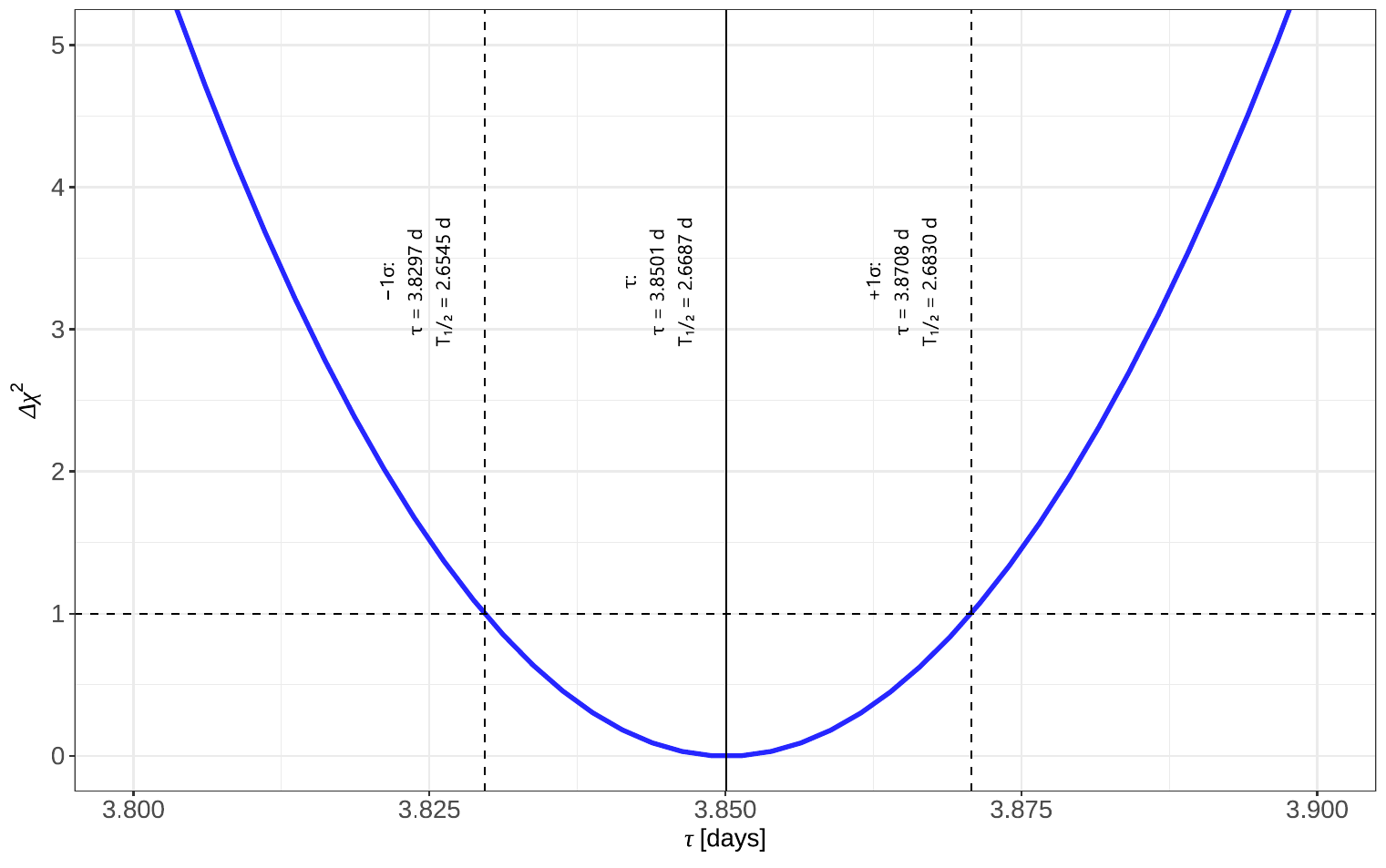}
		\caption{Profile likelihood for the no-offset exponential model, shown as \(\Delta\chi^2\) versus the mean lifetime \(\tau=T_{1/2}/\ln 2\) in days. The solid vertical line marks the best-fit value, while the dashed vertical lines mark the approximate one-standard-deviation interval obtained from the \(\Delta\chi^2=1\) crossing for one parameter of interest. The horizontal dashed line at \(\Delta\chi^2=1\) indicates the approximate 68.27\% confidence interval threshold under the Gaussian error working model.}
		\label{fig:profile-no-offset}
	\end{figure}

For the no-offset exponential model, the profile in \(T_{1/2}\) gives a best-fit value of the following:
	\[
	T_{1/2}=230{,}575~\mathrm{s}
	=2.6687~\mathrm{d}.
	\]
{Using} the usual one-parameter threshold \(\Delta\chi^2=1\), the approximate \(1\sigma\), or 68.27\%, profile interval for the digitized dataset is as follows:
	\[
	T_{1/2}=[229{,}352,\;231{,}810]~\mathrm{s}
	=[2.6546,\;2.6830]~\mathrm{d}.
	\]
{For} the same no-offset model, the profile in \(A_0\) remains centered on the best-fit value,
\[
	A_0=3.7333~\mathrm{cps}.
\]
Figure~\ref{fig:profile-no-offset} shows the \(\Delta\chi^2\) profile expressed in terms of the mean lifetime,
	\[
	\tau=T_{1/2}/\ln 2.
	\]
{The} best-fit value corresponds to
	\[
	\tau=3.8501~\mathrm{d},
	\qquad
	T_{1/2}=\tau\ln2=2.6687~\mathrm{d}.
	\]
	\vspace{-21pt}
	
The profile representation makes the objective function’s curvature explicit and provides a transparent description of uncertainty when fitted parameters are strongly coupled. In this case, the profile interval is obtained by refitting the nuisance parameter at each fixed value of \(T_{1/2}\), rather than by relying only on the local linearized covariance matrix.

{The profile interval and the covariance uncertainty were initially calculated using different uncertainty scales. The $\Delta\chi^2=1$ profile gives limits of $-0.0141$ and $+0.0143~\mathrm{d}$, which agree with the unscaled covariance uncertainty of about $0.0142~\mathrm{d}$. However, the reported covariance uncertainty of $0.0171~\mathrm{d}$ includes the scale factor $s=1.201$. Applying a scale factor when the observed scatter is larger than expected follows the uncertainty-scaling practice recommended by the Particle Data Group (PDG)~\cite{Workman2022PDG}. Erler and Ferro-Hernández discuss the purpose of PDG scale factors and possible alternatives in detail~\cite{ErlerFerroHernandez2020}. When the same scale factor is applied to the profile interval, the limits become approximately $-0.0169$ and $+0.0172~\mathrm{d}$, giving the interval $[2.6518,2.6859]~\mathrm{d}$. Therefore, the covariance and profile results agree when the same uncertainty scale is used.}

	\subsection{Auxiliary Nonnegative Background Check}

Although the digitized dataset was already background-subtracted, a nonnegative background term was used as a conservative stress test. The check did not move the best-fit half-life: the profile minimum remained at the no-offset value, \(T_{1/2}=2.6687~\mathrm{d}\), within the reported precision. The main information obtained from this test was not a new half-life but the appearance of boundary behavior. When the fitted nonnegative background reaches zero, the likelihood becomes non-regular and ordinary \(\Delta\chi^2\) interval rules are not appropriate. The details are provided in Appendix~\ref{app:auxiliary_checks}.
	
For the processed net-count rate dataset used here, the signed residual offset profile in Section~\ref{sec:residual_offset} is the more relevant test of baseline sensitivity.

	\subsection{Sensitivity to Analysis-Level Perturbations}
	
The most consequential robustness question for digitized decay data is how sensitive \(T_{1/2}\) is to small analysis-level perturbations that arise when raw spectra and original subtraction details are unavailable. Therefore, we applied controlled perturbations and refitted the no-offset model in Equation~\eqref{eq:baseline}.
	
{We shifted all time values by up to $\pm60~\mathrm{s}$ and found no meaningful change in the fitted half-life. The largest numerical change was about $6.4\times10^{-5}~\mathrm{s}$. This result agrees with the expected behavior of a single-exponential fit with a free normalization parameter. In contrast, increasing the time scale by $0.5\%$ increased the fitted half-life by approximately the same percentage, or about $1153~\mathrm{s}$ ($0.0133~\mathrm{d}$).}
	
\begin{figure}[h]
		\includegraphics[width=0.99\linewidth]{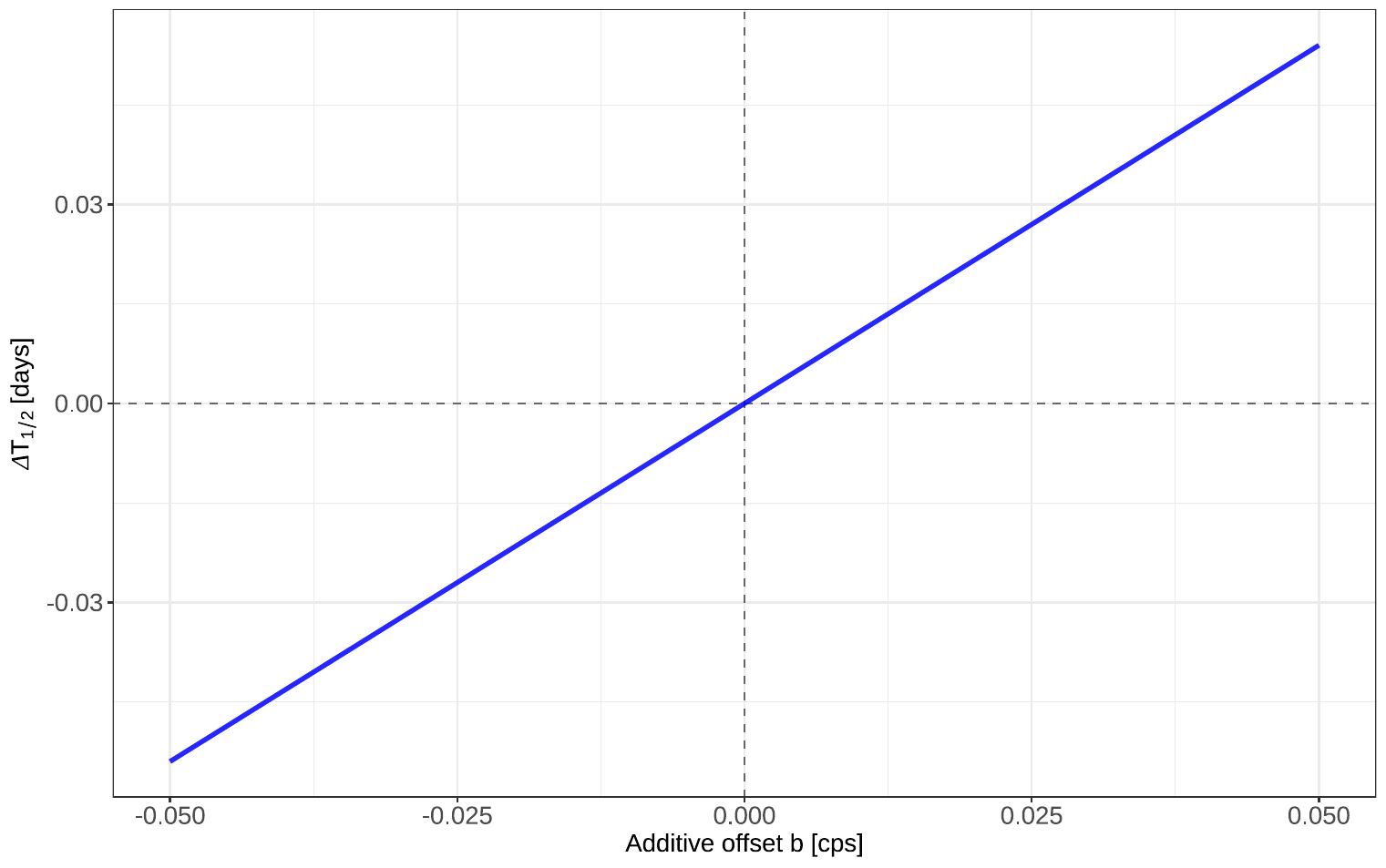}
		\caption{{Sensitivity}  of the fitted half-life to a uniform additive offset \(b\) applied to all digitized count rates. The offset scan shows that baseline-like shifts at the level of tens of milli-cps can move the inferred \(T_{1/2}\) by up to about \(0.0540~\mathrm{d}\), or \(1.30~\mathrm{h}\). This makes the additive baseline offset the dominant analysis-level perturbation in this figure-only reconstruction.}
		\label{fig:offset-sensitivity}
	\end{figure}

In contrast, adding a uniform offset \(b\) to all digitized count rates, \(A_i\to A_i+b\), over the range \(-0.05\le b\le0.05~\mathrm{cps}\), shifts the fitted half-life by up to
	\[
	|\Delta T_{1/2}|=4666.7~\mathrm{s}=0.0540~\mathrm{d},
	\]
or about \(1.30~\mathrm{h}\). \textls[-15]{Figure~\ref{fig:offset-sensitivity} shows the dominant sensitivity. This result suggests that baseline-like distortions can control the robustness envelope when only figure-level data are available.}
	
Influence checks lead to a consistent picture. Leave-one-out refits can change the inferred half-life by up to
\[
0.00731~\mathrm{d},
\]
or about \(10.5~\mathrm{min}\). Removing early time points produces larger shifts than removing late time points. For example, dropping the earliest \(10\%\) of the points shifts the half-life by the~following:
	\[
	\Delta T_{1/2}=-2385~\mathrm{s}=-0.0276~\mathrm{d},
	\]
dropping the latest \(10\%\) shifts it by
	\[
	\Delta T_{1/2}=+424~\mathrm{s}=+0.00491~\mathrm{d}.
	\]
{This} {asymmetry is expected because the early points strongly constrain the normalization and, through the strong scale--half-life correlation, influence the inferred decay~constant.}
	
	\subsection{Residual Offset Nuisance Profiling for Background-Subtracted Data}
	\label{sec:residual_offset}
	
This subsection examines whether a small signed baseline offset can trade off against the fitted decay constant. This is relevant because the published plotted dataset, including peak extraction and background subtraction, had already been processed in the source study. Therefore, an added constant in the digitized dataset should not be interpreted as a new physical background rate. It is better treated as a possible residual offset from imperfect subtraction, digitization, or small peak area reconstruction effects.
	
To test this sensitivity, we used the signed offset model in Equation~\eqref{eq:offmodel}, treating \(B_{\mathrm{off}}\) as a nuisance parameter. For each fixed \(T_{1/2}\), the model is linear in \((A_0,B_{\mathrm{off}})\), so weighted least squares (LS) can solve these two nuisance parameters exactly. This gives a smooth profile in \(T_{1/2}\) and avoids the branch-sticking behavior that can occur in constrained nonlinear~scans.
	
The no-offset reference fit gives
	\[
	T_{1/2}=230{,}575~\mathrm{s}=2.6687~\mathrm{d},
	\qquad
	A_0=3.7333~\mathrm{cps}.
	\]
{The} weighted mean residual is close to zero.
	\[
	B_0=-1.12\times10^{-4}~\mathrm{cps},
	\]
	where the median point uncertainty and weighted residual root mean square (RMS) are
	\[
	\sigma_{\mathrm{point}}=0.02862~\mathrm{cps},
	\qquad
	\sigma_{\mathrm{res,rms}}=0.03301~\mathrm{cps}.
	\]
{These} numbers indicate that the no-offset residuals are centered very close to zero. However, they do not rule out a shallow trade-off between a constant offset and the decay constant over the finite observation window.
	
Two offset profile variants were considered. In the first, denoted ``LS: free'', \(B_{\mathrm{off}}\) is unconstrained. In the second, denoted ``LS: prior'', a weak Gaussian penalty is applied according to Equation~\eqref{eq:offsetprior}. The penalty is centered at \(B_0\). Its width is chosen conservatively from the free-offset profile: the empirical width near \(\Delta\chi^2\le1\) is
	\[
	\sigma_{B,\mathrm{emp}}=0.0976~\mathrm{cps},
	\]
and the final softened width used in the scan is as follows:
	\[
	\sigma_B=0.2928~\mathrm{cps}.
	\]
	
Allowing a signed residual offset opens a shallow likelihood valley. For the unconstrained offset model, the minimum profile occurs near
	\[
	T_{1/2}\approx2.85~\mathrm{d},
	\qquad
	\hat{B}_{\mathrm{off}}\approx -0.167~\mathrm{cps},
	\]
with \(A_0\approx3.89\) cps. With weak Gaussian regularization, the minimum remains similar to
	\[
	T_{1/2}\approx2.82~\mathrm{d},
	\qquad
	\hat{B}_{\mathrm{off}}\approx -0.137~\mathrm{cps},
	\]
with \(A_0\approx3.86\) cps. These offsets are larger than the median digitized point uncertainty. Therefore, they are not interpreted as measured residual backgrounds. Instead, they mark the direction in the parameter space where a constant baseline term can compensate for a change in the decay constant.
	
Compared with the no-offset fit, the signed offset profiles move from the minimum to larger half-lives. The weak-prior model gives a minimum near \(T_{1/2}\approx2.82~\mathrm{d}\), which is about \(0.151~\mathrm{d}\), or \(3.6~\mathrm{h}\), above the no-offset value. This corresponds to a relative shift of about \(5.7\%\). The unconstrained model gives a minimum near \(T_{1/2}\approx2.85~\mathrm{d}\), which is about \(0.181~\mathrm{d}\), or \(4.4~\mathrm{h}\), above the no-offset value. This corresponds to a relative shift of about~\(6.8\%\).

These shifted minima should not be treated as revised half-life estimates. They show that a constant baseline-like offset can partly mimic a change in the decay rate over the limited time window of the digitized dataset. The size of this diagnostic shift is notable because it is comparable to the difference between the individual room-temperature and \(12~\mathrm{K}\) fits reported in the original publication, \(2.669~\mathrm{d}\) and \(2.817~\mathrm{d}\), respectively~\cite{spillane2007}. It is also larger than the difference between the reported weighted averages, \(2.706~\mathrm{d}\) at room temperature and \(2.802~\mathrm{d}\) at \(12~\mathrm{K}\). This comparison does not show that a baseline offset caused the temperature-dependent difference in the original experiment. The comparison shows only that a baseline-like effect of this scale would be large enough to affect such a comparison. Without the original spectra, peak fits, background subtraction details, or detector model, the digitized dataset alone cannot separate this baseline-like offset from a change in the decay constant.
	
\begin{figure}[h]
		\includegraphics[width=0.99\linewidth]{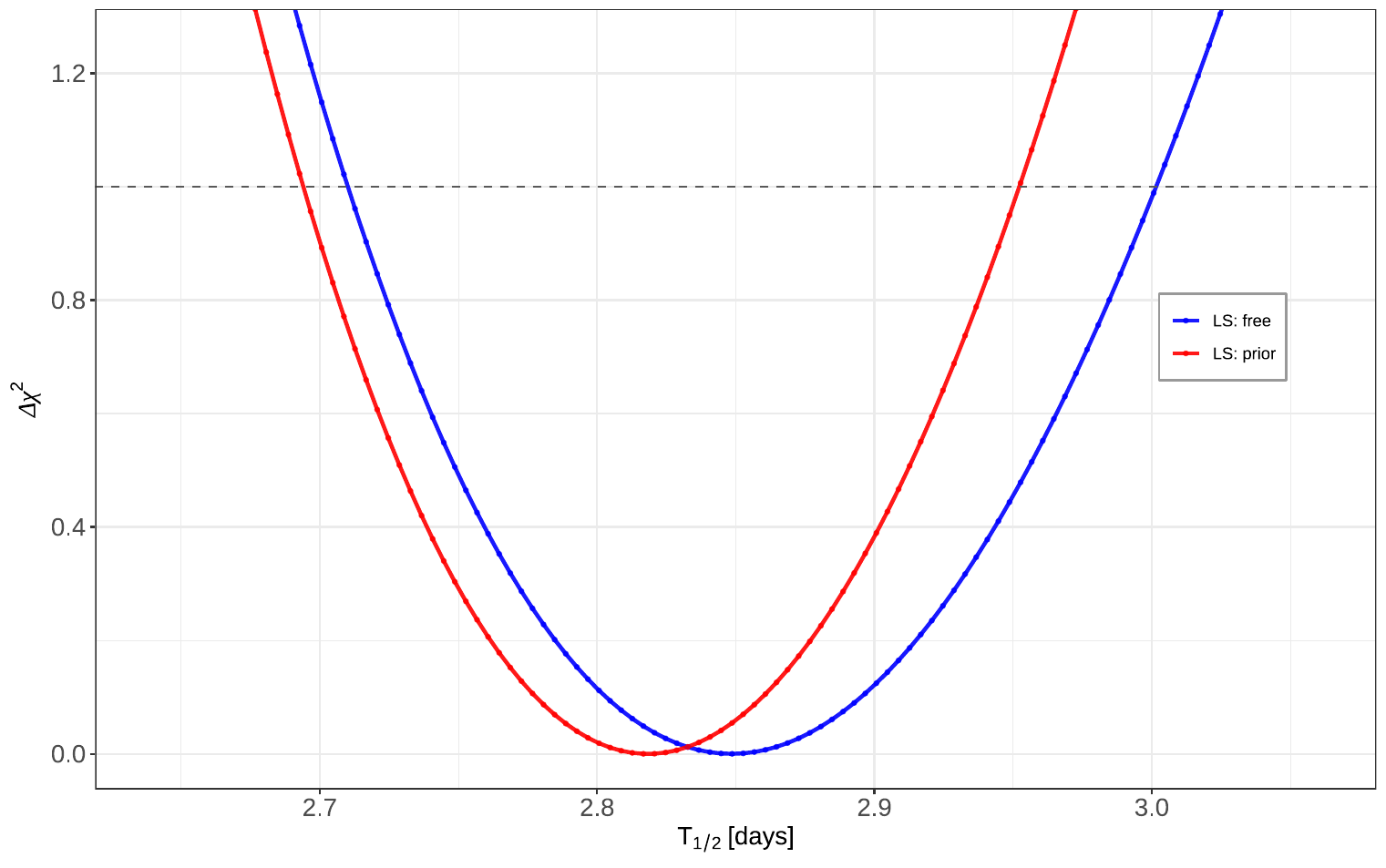}
		\caption{{Residual}  offset profile for the background-subtracted, digitized figure-level dataset. The curves show \(\Delta\chi^2(T_{1/2})\) for the signed residual offset model with a free offset (LS: free) and a weakly constrained offset (LS: prior). The dashed horizontal line marks \(\Delta\chi^2=1\), which is used to identify the approximate one-standard-deviation interval for \(T_{1/2}\). The broad curves show that the fitted baseline offset and decay constant can compensate for each other over the limited observation period. Therefore, this profile is used to assess uncertainty in the baseline offset, not as the primary half-life estimate.}
		\label{fig:boff-profile}
	\end{figure}

	Figure~\ref{fig:boff-profile} shows the corresponding \(\Delta\chi^2\) profiles. The no-offset regression corresponds to the special case \(B_{\mathrm{off}}=0\). Once \(B_{\mathrm{off}}\) is varied, the profile exposes a broad direction in \((T_{1/2},B_{\mathrm{off}})\) space. The residual offset profile is therefore diagnostic rather than prescriptive: it quantifies how strongly baseline-like distortions can affect a figure-only reconstruction, but it is not used as the primary half-life estimate.

{A formal comparison of the no-offset, signed offset, and regularized signed offset models is outside the scope of this study. The signed offset model is used only to show how a possible baseline shift could affect the result. It is not treated as a separate physical model. A detailed comparison of other models, including correlated digitization errors, uncertain extra parameters, and regularization, could be completed in future work.}
	
\textls[-15]{A complementary Fourier transform residual baseline diagnostic is given in \mbox{Appendix~\ref{app:fft_baseline}}. }This diagnostic estimates \(B_{\mathrm{FFT}}\) from the first nonzero Fourier harmonic and provides a separate scale check for the residual offset interpretation used in this study.

	\subsection{Sensitivity Envelope for Figure-Level Reconstruction}
	
	Table~\ref{tab:sensitivity-summary} summarizes the main sensitivity diagnostics at the analysis level. The entries are diagnostic shifts relative to the no-offset fit, not a formal systematic uncertainty budget. Some tests represent plausible figure-level effects, such as time-origin shifts, time-scale distortions, and uniform count rate offsets. Other tests, such as leave-one-out refits and early/late truncation, probe which parts of the digitized dataset have the most leverage. Because the original spectra and analysis logs are unavailable, these effects cannot be uniquely separated into digitization effects and the properties of the processed source data.
	
	\vspace{-6pt}
	\begin{table}[H]

\caption{{Summary}  of analysis-level sensitivity diagnostics for the no-offset fit. The listed changes are diagnostic shifts relative to the baseline fit, not a formal systematic uncertainty budget.}
		\label{tab:sensitivity-summary}
		\small
		
\begin{tabularx}{\textwidth} {lLLL}
			\toprule
\textbf{Component or Diagnostic} & \boldmath{\(T_{1/2}\) \textbf{Shift [s]}} & \boldmath{\(T_{1/2}\) \textbf{Shift [d]}} & \textbf{Relative [\%]} \\
			\midrule
Central value from no-offset fit & 230{,}575 & \(2.6687\) & -- \\
Covariance matrix standard error & \(\pm1477\) & \(\pm0.0171\) & \(\pm0.641\) \\
Profile interval, \(\Delta\chi^2=1\) & \(-1223,\,+1235\) & \(-0.0142,\,+0.0143\) & \(-0.530,\,+0.536\) \\
Absolute time shift, up to \(\pm60~\mathrm{s}\) & \(\le\)6.4\(\times10^{-5}\) & \(\le\)7.4\(\times10^{-10}\) & negligible \\
Time-scale increase, \(+0.5\%\) & \(+1153\) & \(+0.0133\) & \(+0.500\) \\
Uniform offset, \(-0.05 \le b \le 0.05~\mathrm{cps}\) & \(4666.7\) & \(0.0540\) & \(2.02\) \\
Constant uncertainty check, \(\sigma=\mathrm{median}(\sigma_i)\) & \(480.0\) & \(0.00555\) & \(0.208\) \\
Leave-one-out maximum & \(631.3\) & \(0.007307\) & \(0.274\) \\
			Drop earliest \(10\%\) of points & \(-2385\) & \(-0.0276\) & \(-1.03\) \\
			Drop latest \(10\%\) of points & \(+424\) & \(+0.00491\) & \(+0.184\) \\
			\bottomrule
		\end{tabularx}
		
	\end{table}
	
Table~\ref{tab:sensitivity-summary} defines a practical sensitivity envelope for figure-level reconstruction. The entries are diagnostic shifts relative to the no-offset fit, not components of a formal systematic uncertainty budget.
	
A uniform count rate offset means that every digitized point is shifted by the same amount,
	\[
	A_i \rightarrow A_i+b,
	\]
where \(A_i\) is the digitized count rate at time \(t_i\), and \(b\) is measured in cps. A uniform count rate offset means a constant offset applied equally to all digitized points, not a random point-by-point perturbation. For the digitized data, this offset represents possible figure-level effects, such as baseline placement, graphical extraction bias, or a small residual offset inherited from the published background-subtracted curve.
	
The largest tested effect is the uniform count rate offset, which changes the fitted half-life by up to
	\[
	|\Delta T_{1/2}|=4666.7~\mathrm{s}=0.0540~\mathrm{d}.
	\]
{This} value is interpreted as a conservative figure-level sensitivity scale, not as a calibrated systematic uncertainty of the original Spillane et al. measurement~\cite{spillane2007}. It is nevertheless useful for judging the scale. The original publication reported a room-temperature weighted average of \(2.706~\mathrm{d}\) and a \(12~\mathrm{K}\) weighted average of \(2.802~\mathrm{d}\), a difference of about \(0.096~\mathrm{d}\). A baseline-like systematic effect of the size tested here would be large enough to affect comparisons between temperature runs. However, this figure-level analysis cannot determine whether such an effect existed in the original raw data. It only shows that the half-life estimates from this time window are sensitive to small baseline-like~offsets.
	
If these perturbations were applied within the original raw data analysis, some of them could be interpreted as systematic uncertainty components. For example, a time-scale distortion would correspond to a timing-calibration effect, and a controlled baseline variation could be related to peak area extraction or continuum subtraction. In the present work, however, only digitized points from a published figure are available. Therefore, the entries in Table~\ref{tab:sensitivity-summary} are not combined into a total systematic uncertainty.
	
As another check on the digitization process, we performed a point-picking jitter test. In this test, the digitized time and count rate points were moved slightly at random, and the no-offset exponential model was refit each time. This test determines whether the half-life result strongly depends on the graphical placement of the digitized points. We generated 500 jittered datasets, and all 500 fits converged. The fitted half-lives had a median of \(T_{1/2}=2.6688~\mathrm{d}\), a mean of \(T_{1/2}=2.6693~\mathrm{d}\), and a standard deviation of \(0.0145~\mathrm{d}\), corresponding to about \(0.54\%\) of the no-offset half-life. The one-sigma percentile interval was \([2.6550,2.6835]~\mathrm{d}\), or approximately \(-0.51\%\) to \(+0.55\%\) relative to the no-offset value. This interval is close to the no-offset \(\Delta\chi^2=1\) profile interval, \([2.6545,2.6830]~\mathrm{d}\). This indicates that reasonable point-picking changes are not the main source of figure-level sensitivity for this dataset.

{The jitter distribution is interpreted as a repeatability and sensitivity diagnostic, not as a calibrated one-standard-deviation uncertainty. It does not include the full common-axis calibration covariance, and it cannot be propagated as though it were independent of the extracted error bars.}

The signed residual offset profile discussed in Section~\ref{sec:residual_offset} is related to this sensitivity question but is not included in Table~\ref{tab:sensitivity-summary}. The reason is that it is not a bounded perturbation of the no-offset fit. Instead, it is an alternative nuisance parameter model. It gives a larger apparent shift, moving the profile minimum to \(T_{1/2}\approx2.82\)--\(2.85~\mathrm{d}\), or about \(5.7\)--\(6.8\%\) above the no-offset value. Therefore, this shifted minimum is treated as a diagnostic of non-identifiability: it shows that a baseline-like offset and the decay constant are not separable from the digitized data alone. The Fourier transform check in Appendix~\ref{app:fft_baseline} provides a separate residual baseline scale estimate and supports the interpretation that these offset tests are diagnostics rather than physical background measurements.
	
With this convention, the no-offset result can be summarized with a separate figure-level sensitivity scale as
	\[
	T_{1/2}=(230{,}575 \pm 1{,}477_{\mathrm{stat}} \pm 4{,}667_{\mathrm{fig}})~\mathrm{s}
	=(2.6687 \pm 0.0171_{\mathrm{stat}} \pm 0.0540_{\mathrm{fig}})~\mathrm{d},
	\]
where ``stat'' denotes the approximate one-standard-deviation statistical uncertainty from the weighted fit, and ``fig'' denotes the figure-level sensitivity scale from the additive-offset scan. The ``fig'' term is not a calibrated 68.27\% confidence interval. It is a diagnostic scale that shows how much the fitted half-life can move under the tested figure-level offset perturbation.
	
	\subsection{Pairwise Ratio and MFV Robustness Diagnostics}
	
As a complementary robustness check, we applied the pairwise ratio and MFV workflow to the same digitized \Au{} dataset. The pairwise ratio method removes the fitted normalization parameter by comparing the count rate ratios between pairs of points. This is useful because the regression fit estimates \(A_0\) and \(\tau\) simultaneously, resulting in a strong \((A_0,\tau)\) correlation.
	
The pairwise median and MFV uncertainties are reported as nonparametric percentile bootstrap intervals. In each bootstrap replicate, the original digitized time--count rate points were resampled with replacement, sorted by time, and the all-valid pairwise lifetime distribution was recomputed. The reported 68.27\% intervals are the 15.865-th and 84.135-th percentiles of the resulting bootstrap distribution, corresponding to the central 68.27\% percentile interval. This convention follows the percentile bootstrap construction, in which the ordered bootstrap estimates themselves define the confidence limits~\cite{Puth2015BootstrapCI}.
	
This follows the standard percentile bootstrap idea of estimating the sampling distribution by resampling the observed data~\cite{EfronTibshirani1993,DavisonHinkley1997Bootstrap}. Practical guidance for resampling-based confidence limits recommends having at least about 10--15 usable observations, with more observations preferred~\cite{Singh2006EPAUCL}. For datasets with fewer than about ten observations, hybrid parametric bootstrapping (HPB) can be used as an alternative diagnostic approach when individual measurement uncertainties are available, because HPB incorporates the uncertainty associated with each data point into the resampling procedure~\cite{Golovko2025}. The digitized \Au{} data set contains substantially more than this minimum, but the intervals are still used here only as robustness diagnostics.
	
The pairwise median gives
	\[
	\tau_{\mathrm{med}}=3.8355^{+0.0282}_{-0.0293}~\mathrm{d},
	\qquad
	T_{1/2,\mathrm{med}}=2.6586^{+0.0195}_{-0.0203}~\mathrm{d},
	\]
where the uncertainties are bootstrap 68.27\% intervals. The pairwise MFV gives
	\[
	\tau_{\mathrm{MFV}}=3.8276^{+0.0282}_{-0.0279}~\mathrm{d},
	\qquad
	T_{1/2,\mathrm{MFV}}=2.6531^{+0.0195}_{-0.0193}~\mathrm{d}.
	\]
{Both} pairwise central values are lower than the no-offset regression result.
	\[
	\tau_{\mathrm{reg}}=3.8501^{+0.0247}_{-0.0247}~\mathrm{d},
	\qquad
	T_{1/2,\mathrm{reg}}=2.6687^{+0.0171}_{-0.0171}~\mathrm{d}.
	\]
	
{Table~\ref{tab:estimator-comparison} collects the estimator comparison in one place. The pairwise differences are small compared with the half-life itself, and the diagnostic intervals overlap; however, the pairwise estimates are correlated because they reuse the same 78 observations.}


\begin{table}[H]  
 \caption{Comparison of the primary regression estimate and supporting pairwise diagnostics. Pairwise intervals are 68.27\% percentile bootstrap intervals. The regression interval is the reported scale-adjusted local covariance interval.} \label{tab:estimator-comparison} \small \begin{tabular}{@{}p{0.22\linewidth} >{\centering\arraybackslash}p{0.14\linewidth} >{\centering\arraybackslash}p{0.22\linewidth} >{\centering\arraybackslash}p{0.32\linewidth}@{}} 
\toprule 
\textbf{Estimator} &{$\bm {T_{1/2}}$}  \textbf{[d]} & \textbf{68.27\% Interval [d]} & \textbf{Difference from Regression [d]} \\ 
\midrule 
No-offset regression & 2.6687 & $[2.6516,2.6858]$ & -- \\ Pairwise median & 2.6586 & $[2.6383,2.6781]$ & $-0.0101$ \\ Pairwise MFV & 2.6531 & $[2.6338,2.6726]$ & $-0.0156$ \\ 
\bottomrule 
	\end{tabular} 
\end{table}

The regression uncertainties are local covariance matrix statistical uncertainties. The scale factor defined in Equation~(\ref{eq:chi2}) is reported separately because Particle Data Group practice allows fit uncertainties to be enlarged when the reduced chi-square is greater than unity~\cite{Workman2022PDG}. This distinction is important because the digitized points scatter more around the fitted curve than their assigned point uncertainties predict. In the original Figure~2 fit, Spillane et al. reported a goodness-of-fit value close to unity, \(\chi^2_\nu\simeq1.06\)~\cite{spillane2007}, corresponding to \(s\simeq1.03\). The digitized reconstruction gives a larger value, \(s\simeq1.20\), indicating that the reconstructed point uncertainties do not capture all scatter sources, such as graphical extraction uncertainty and possible correlations among digitized points.

\begin{figure}[h]
	\includegraphics[width=0.99\linewidth]{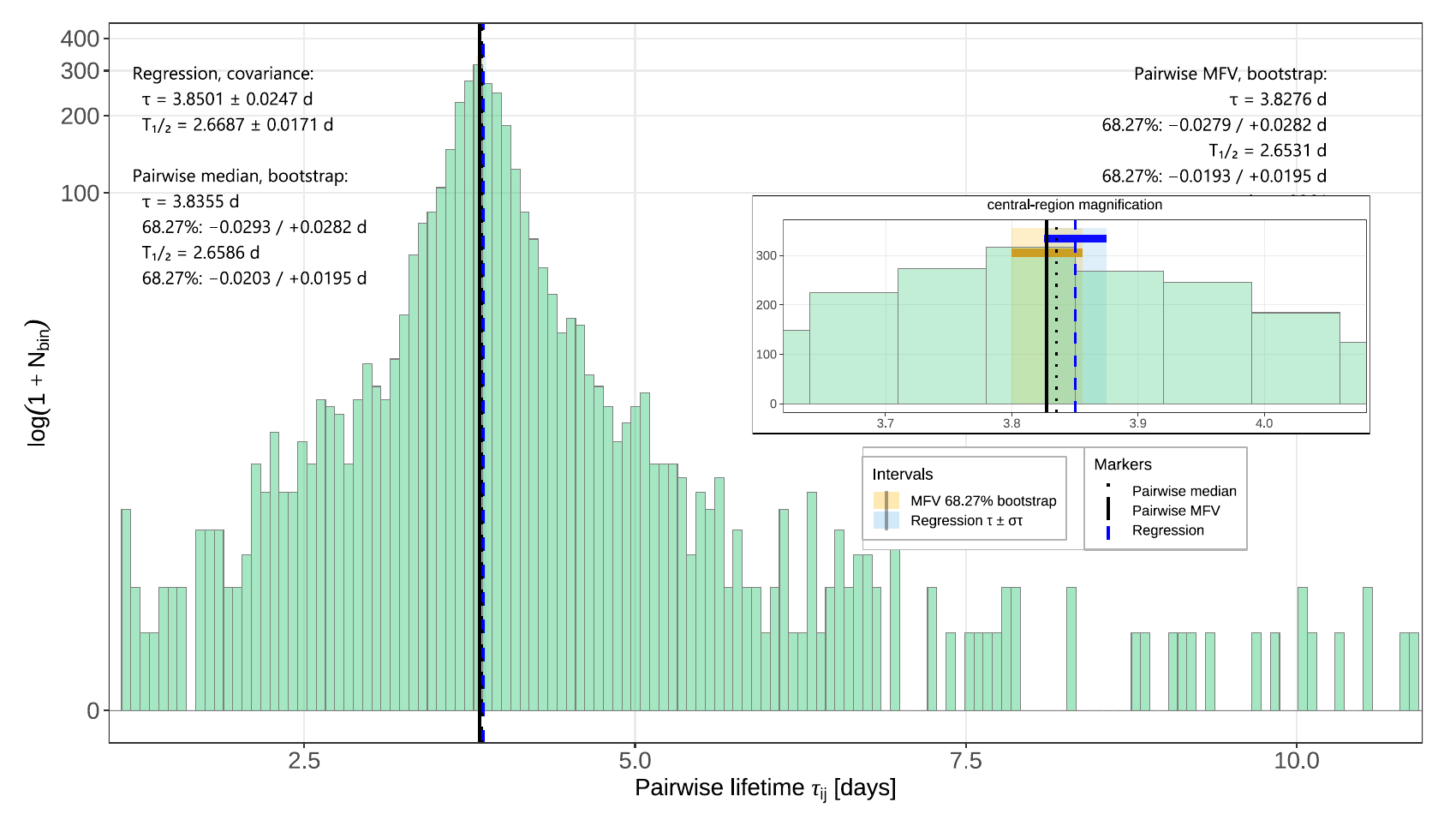}
	\caption{{All}-valid-pair distribution of pairwise lifetime estimates, \(\tau_{ij}\), from the digitized \Au{} decay data. The green histogram shows the pairwise lifetime distribution. Vertical markers show the regression estimate, pairwise median, and pairwise MFV. The blue band shows the regression covariance interval, \(\tau\pm\sigma_{\tau}\), and the gold band shows the 68.27\% bootstrap interval for the pairwise MFV. The inset magnifies the central estimator region. The overlap between the regression and MFV intervals indicates that the ratio-based summaries are consistent with the regression result within their diagnostic uncertainty ranges.}
	\label{fig:pairwise-tau-hist}
\end{figure}

Figure~\ref{fig:pairwise-tau-hist} shows why the pairwise summaries are useful but should be interpreted cautiously. The regression estimate, pairwise median, and pairwise MFV all fall in the central high-density region of the pairwise lifetime distribution. The inset shows that the regression covariance interval overlaps the MFV bootstrap interval. Thus, the ratio-based summaries recover the same lifetime scale as the regression fit within their diagnostic uncertainty ranges. Because the pairwise distribution is long-tailed, robust summaries such as the median and MFV are preferable to a simple mean for describing its central region. These summaries reduce the influence of extreme pairwise values, but they remain diagnostic quantities because many pairwise ratios share the same original digitized observations.

Close-time pairs can make the pairwise estimator unstable because the expected decay between nearby points is comparable to the measurement and digitization scatter. In this case, a later point can fluctuate above an earlier point, or the logarithmic count rate ratio can become too small to provide a stable lifetime estimate. To test this effect, the MFV analysis was repeated with minimum time-separation cuts applied to the all-valid pair set. For the digitized dataset, the all-valid-pair MFV result is \(T_{1/2,\mathrm{MFV}}=2.6531~\mathrm{d}\). Requiring \(\Delta t\ge12~\mathrm{h}\) gives \(T_{1/2,\mathrm{MFV}}=2.6583~\mathrm{d}\), an upward shift of \(0.0052~\mathrm{d}\). This indicates that close pairs contribute to the downward displacement of the pairwise summary, but they do not remove it. Therefore, the remaining difference is treated as part of the finite-window behavior of the pairwise estimator.
	
As an additional ratio-based smoothing check, we applied the mean displaced ratio (MDR) diagnostic~\cite{DysonIsenberg1971MDR} (details are given in Appendix~\ref{app:mdr}). Because the published \Au{} dataset was already background-subtracted, the MDR transformation was applied to net count rate data rather than to raw count rates with an unknown physical background. Using the Dyson--Isenberg distinct indexing convention, the MDR fits gave half-lives within \(-0.43\%\) to \(+1.36\%\) of the no-offset regression value. The MDR points are correlated because they share the same digitized observations. Thus, the fitted uncertainties are only interpreted diagnostically. Therefore, the MDR results support the same half-life scale as the regression and pairwise ratio analyses, but they are not used as preferred half-life~estimates.
	
	\subsection{Toy Monte Carlo Diagnostic with an Idealized Time Grid}
	
In this subsection, idealized synthetic data are used to separate estimator behavior from digitization effects. As a control test, we applied the same regression, profiling, pairwise-median, and MFV workflow to a synthetic decay dataset. The toy Monte Carlo dataset contains 78 points on a uniform \(3600~\mathrm{s}\) time grid. This matches the nominal one-hour sampling of the source experiment, but it does not include the small nonuniformities introduced by digitizing the published figure. Toy count rates were generated using the no-offset exponential model with point uncertainties comparable to those used in figure-level reconstruction. This test aims to check whether the analysis pipeline behaves as expected when the model assumptions are satisfied under regular sampling.
	
For the toy dataset, the no-offset weighted regression gives the following:
	\[
	A_{0,\mathrm{toy}}=(3.6808\pm0.0153)~\mathrm{cps},
	\qquad
	T_{1/2,\mathrm{reg,toy}}=(230558.5\pm2241)~\mathrm{s}
	=(2.6685\pm0.0259)~\mathrm{d}.
	\]
{Fitted} normalization and half-life are strongly correlated,
	\[
	\rho(A_0,T_{1/2})=-0.828.
	\]
{This} is expected for a two-parameter exponential fit over a finite observation window. The goodness-of-fit is close to the ideal,
	\[
	\chi^2/\mathrm{ndf}=74.53/76=0.981,
	\qquad
	s=\sqrt{\chi^2/\mathrm{ndf}}=0.990.
	\]
{Thus}, the assigned point uncertainties for the toy data are consistent with the observed residual scatter, and no uncertainty inflation is indicated.
	
The toy central values are close to the published Figure~2 fit of {Spillane et al.}~\cite{spillane2007},
	\[
	A(0)=(3.68\pm0.04)~\mathrm{cps},
	\qquad
	T_{1/2}=(2.669\pm0.017)~\mathrm{d},
	\]
{for} which the reported goodness-of-fit value was also close to unity, \(\chi^2_\nu\simeq1.06\). The toy calculation does not reproduce the published statistical uncertainty exactly, and this is expected. The toy data are an idealized count rate curve and not a reconstruction of the full original analysis chain. The uncertainty reported in the source study could depend on how the peak areas were extracted, how the background was fitted, how live time and counting statistics were handled, and how covariances and final uncertainties were reported. Because the raw spectra and detailed analysis records are not available, these separate contributions cannot be uniquely reconstructed from the published figure alone.
	
The toy residuals behave as expected for the assumed model-generated data. Of the 78~normalized residuals, 53 fall within \(1\sigma\), 22 fall between \(1\sigma\) and \(2\sigma\), and three fall outside \(2\sigma\). The Shapiro--Wilk, Anderson--Darling, and Lilliefors tests give \(p_{\mathrm{SW}}=0.186\), \(p_{\mathrm{AD}}=0.199\), and \(p_{\mathrm{LF}}=0.507\), respectively, so the residual tests do not reject the Gaussian working model.
	
	As a descriptive comparison with the digitized reconstruction, we also compared the normalized residuals from the toy and digitized fits after subtracting the fitted exponential trend. The two-sample Kolmogorov--Smirnov test was used as a distribution shape diagnostic, and the Wilcoxon--Mann--Whitney rank-sum test was used as a rank/location diagnostic~\cite{Kolmogorov1933,Smirnov1948,Wilcoxon1945,MannWhitney1947}. The corresponding \(p\)-values are \(0.810\) and \(0.970\), respectively. Therefore, these tests do not indicate a clear difference in the residual shape or central location. However, the digitized residuals are broader, with a standard deviation of \(1.19\), compared with \(0.98\) for the toy residuals. This broader residual spread is consistent with the larger scale factor found for the digitized reconstruction.
	
	Table~\ref{tab:toy-control-summary} summarizes the main diagnostics of toy data. The entries are diagnostic shifts relative to the toy no-offset fit, not systematic uncertainties. The table shows that the toy data do not require uncertainty scaling, but they still show the same qualitative finite-window sensitivities as the digitized data. Absolute time shifts are negligible, time-scale distortions scale the fitted half-life directly, and uniform count rate offsets or removal of early points can shift the fitted half-life appreciably.
	
	\begin{table}[H]
		
		\caption{Toy Monte Carlo control diagnostics for no-offset fit. The toy dataset contains 78 points on an idealized uniform \(3600~\mathrm{s}\) time grid. The entries are diagnostic shifts relative to the toy no-offset fit, not systematic uncertainties.}
		\label{tab:toy-control-summary}
		\small
		
\centering 

		\begin{tabularx}{\textwidth} {lLLL}

			\toprule
\textbf{Component or Diagnostic} & \(\bm{T_{1/2}}\) \textbf{Shift [s]} & \textbf{\(\bm{T_{1/2}}\) Shift [d]} &\textbf{ Relative [\%]} \\
			\midrule
Central value from toy no-offset fit & 230{,}558.5 & \(2.6685\) & -- \\
Covariance matrix standard error & \(\pm2241\) & \(\pm0.0259\) & \(\pm0.972\) \\
Profile interval, \(\Delta\chi^2=1\) & \(-2239,\,+2280\) & \(-0.0259,\,+0.0264\) & \(-0.971,\,+0.989\) \\
Absolute time shift, up to \(\pm60~\mathrm{s}\) & \(\le\)1.4\(\times10^{-4}\) & \(\le\)1.6\(\times10^{-9}\) & negligible \\
Time-scale increase, \(+0.5\%\) & \(+1153\) & \(+0.0133\) & \(+0.500\) \\
Uniform offset, \(-0.05 \le b \le 0.05~\mathrm{cps}\) & \(4708.3\) & \(0.0545\) & \(2.04\) \\
Constant uncertainty check, \(\sigma=\mathrm{median}(\sigma_i)\) & \(853.9\) & \(0.00988\) & \(0.370\) \\
Leave-one-out maximum & \(1102\) & \(0.01276\) & \(0.478\) \\
			Drop earliest \(10\%\) of points & \(-3928\) & \(-0.0455\) & \(-1.70\) \\
			Drop latest \(10\%\) of points & \(+850\) & \(+0.00984\) & \(+0.369\) \\
			\bottomrule
			
		\end{tabularx}
		
	\end{table}
	
Toy control helps separate the two effects. First, the regression workflow behaves normally when the data are generated from the assumed exponential model: the fitted half-life is recovered and \(s\simeq1\). Second, some sensitivities remain even in clean toy data. In particular, the signed offset profile is broad when \(B_{\mathrm{off}}\) is varied. This means that the offset--lifetime trade-off is partly a finite-window identifiability problem and not only a feature of the digitized dataset. The signed offset profile should therefore be interpreted as a diagnostic of sensitivity to baseline-like distortions and not as evidence for a measured residual background.
	
The toy data were also processed using the pairwise ratio and MFV workflow. The pairwise median gives
	\[
	\tau_{\mathrm{med,toy}}=3.8180^{+0.0480}_{-0.0379}~\mathrm{d},
	\qquad
	T_{1/2,\mathrm{med,toy}}=2.6464^{+0.0333}_{-0.0263}~\mathrm{d},
	\]
where the uncertainties are bootstrap 68.27\% intervals. The pairwise MFV gives
	\[
	\tau_{\mathrm{MFV,toy}}=3.8043^{+0.0428}_{-0.0413}~\mathrm{d},
	\qquad
	T_{1/2,\mathrm{MFV,toy}}=2.6369^{+0.0297}_{-0.0286}~\mathrm{d}.
	\]
{Both} pairwise central values are lower than the result of the toy regression,
	\[
	\tau_{\mathrm{reg,toy}}=3.8498^{+0.0374}_{-0.0374}~\mathrm{d},
	\qquad
	T_{1/2,\mathrm{reg,toy}}=2.6685^{+0.0259}_{-0.0259}~\mathrm{d}.
	\]	
{However}, the 68.27\% intervals still overlap.
This result indicates that a lower pairwise median or MFV central value can occur even when the data are generated from a clean no-offset exponential model. Therefore, the downward shift is not, by itself, evidence for a different physical half-life. It reflects the finite observation window and the shared-point structure of the pairwise ratios. The bootstrap intervals capture this additional estimator spread, and the toy regression, pairwise median, and pairwise MFV results remain mutually consistent within their estimated uncertainties.
	
An extended toy diagnostic with three times the observation duration is provided in Appendix~\ref{app:extended_toy_control}. The longer observation window reduces the normalization--lifetime correlation, weakens the signed offset degeneracy, and moves the pairwise median and MFV estimates closer to the regression result. This supports the interpretation that part of the pairwise/MFV shift and much of the offset--lifetime behavior come from the finite observation window rather than from a different physical half-life.

\section{Discussion and Future Work}
\label{sec:discussion}

The main purpose of this study was not to remeasure the \Au{} half-life from the original experimental records. Those records include the raw spectra, peak integration details, live-time corrections, detector stability information, and background fits, none of which are available from the published figure alone. Instead, the study asks what can still be learned from reduced radiation sensor data when only a published plot is available. The specific test was whether the published room-temperature decay curve result can be reproduced from digitized points and which parts of the inference become sensitive to figure-level effects, such as baseline treatment, time-window length, uncertainty assignment, and estimator choice. In this sense, the \Au{} curve is used as a case study for reduced-data quality assurance in radiation sensor analysis, rather than as a new measurement of the half-life.

The no-offset exponential regression reproduces the published half-life scale. This is an important consistency check, but it does not by itself prove that the reconstruction is fully robust. The fit estimates the normalization and lifetime at the same time, and these two parameters are strongly correlated. Over the finite time window of the digitized dataset, changes in the early points, baseline level, or uncertainty model can shift the fitted half-life while leaving the fitted model visually acceptable.

The profile likelihood analysis is useful because it checks the shape of the fit objective beyond the local covariance approximation. A compact and nearly parabolic profile supports ordinary interval reporting. A broad, flat, or irregular profile indicates that the parameter is weakly constrained or that nuisance parameters can partly compensate for changes in the parameter of interest. Profile likelihoods are widely used to construct confidence intervals in the presence of nuisance parameters and to diagnose practical identifiability limits caused by finite or noisy data~\cite{RolkeLopezConrad2005,Raue2009ProfileLikelihood}.

The goodness-of-fit comparison separates behavior caused by the digitized data from behavior caused by the fitting method. The digitized reconstruction gives a scale factor of about \(s\simeq1.20\), whereas the idealized toy Monte Carlo dataset generated from a clean no-offset exponential model gives \(s\simeq0.99\). This indicates that the broader residual scatter in the digitized data is not simply a consequence of the fitting algorithm. The extra scatter is more likely introduced by figure-level effects, such as point extraction, reconstructed error bars, shared axis calibration, finite graphical resolution, or small residual structures already present in the processed curve shown in the publication. Toy simulations and goodness-of-fit checks are commonly used in rare event searches to test statistical behavior and assess background or detector models~\cite{Baxter2021DMReporting}.

The residual comparisons support the same interpretation. After subtracting the fitted exponential trend, the digitized and toy residuals do not show a clear difference in central location or distribution shape according to the Kolmogorov--Smirnov and Wilcoxon--Mann--Whitney diagnostics. The main difference is that the digitized residuals are wider. This is why the scale factor in Equation~(\ref{eq:chi2}) is reported and why a Particle Data Group style enlargement of the statistical uncertainty should be considered when the reduced chi-square is greater than unity~\cite{Workman2022PDG}. The uncertainty should not imply that the digitized points are more precise than their observed scatter supports.

The baseline and offset tests define the main limitation of the figure-level reconstruction. Because the published plotted dataset was already background-subtracted, a nonnegative background term should not be interpreted as a measured physical background. In this setting, a signed residual offset is more appropriate because residual baseline errors from subtraction, digitization, or peak area reconstruction can have either sign. The signed offset profile shows that a constant offset and the decay constant can trade off over the finite observation window. Therefore, the shifted profile minima are not revised half-life estimates and are not measurements of residual background. They show that figure-only data cannot cleanly separate a small baseline-like distortion from a change in half-life without additional experimental information.

The comparison with the low-temperature result should therefore be interpreted cautiously. If a comparable baseline sensitivity were present in the original non-digitized analysis, then a run-dependent analysis effect, such as continuum subtraction, peak area extraction, residual baseline treatment, or normalization stability, could contribute to the apparent room-temperature versus low-temperature difference. This study cannot determine whether such an effect occurred, because the original spectra and analysis records are unavailable.

The sensitivity summary should be interpreted in the same way. The table entries are diagnostic shifts for figure-level reconstruction, not a systematic uncertainty budget for the original Spillane et al. experiment~\cite{spillane2007}. Timing offsets have negligible effect, and a time-scale distortion changes the half-life in the expected proportional way. The largest tested effect is the uniform count rate offset. This is physically reasonable because a constant offset changes the apparent curvature of the decay curve, especially near the low-rate end of the measurement. For this reason, the offset scan is reported as a separate figure-level sensitivity scale, denoted ``fig'', rather than as a calibrated systematic uncertainty of the original measurement.

The pairwise ratio and MFV analyses provide a complementary robustness check because they remove the fitted normalization from each pairwise ratio. This is useful because the regression normalization \(A_0\) and lifetime \(\tau\) are strongly correlated. The pairwise median and pairwise MFV central values are slightly lower than the no-offset regression value, but their bootstrap intervals overlap the regression interval. This means that the ratio-based summaries recover the same lifetime scale within their diagnostic uncertainty ranges. The lower central values should not be interpreted as evidence for another preferred half-life. The pairwise ratios share the same original digitized observations, so they are not independent measurements. The bootstrap intervals are therefore used only as diagnostic resampling intervals, not as a replacement for the regression and profile likelihood results. Bootstrap intervals can also have limitations, especially for small samples, extrema, modes, correlation coefficients, or poorly sampled distribution tails~\cite{Nicholls2014Confidence}.

The toy Monte Carlo controls clarify the meaning of the pairwise shift. The same lower pairwise median and MFV central values appear in idealized toy data generated from a clean no-offset exponential model. This indicates that a downward shift of the pairwise summaries can arise from the finite observation window and from the shared-point structure of the pairwise ratios. It is not, by itself, evidence for a different physical half-life. The extended toy control in Appendix~\ref{app:extended_toy_control} supports this interpretation: when the observation window is lengthened, the normalization--lifetime correlation becomes weaker, the signed offset degeneracy is reduced, and the pairwise median and MFV estimates move closer to the regression result.

The MFV implementation was also checked against the neutron lifetime benchmark of Zhang et al.~\cite{Zhang2022MFVNeutron}. The MFV estimate and percentile bootstrap intervals reproduce the published benchmark within Monte Carlo variation. This validation separates implementation issues from the behavior of the \Au{} digitized dataset. The lower MFV value in the \Au{} pairwise analysis is therefore not an algorithmic failure. It is part of the expected estimator behavior for finite-window exponential data.

Overall, the digitized plotted data points are sufficient to reproduce the broad half-life scale and to test several sensitivity modes. It is not sufficient to reconstruct the original experiment's full uncertainty model, background treatment, detector corrections, or covariance structure. The most defensible result is therefore the no-offset regression half-life, reported with its statistical uncertainty and with a separate figure-level sensitivity scale. The supporting diagnostics point to the same conclusion. The signed offset profile and additive-offset scan show that small baseline-like changes can move the fitted half-life. The pairwise median, MFV, MDR, toy Monte Carlo, and Fourier transform checks recover the same approximate half-life scale but show that small estimator shifts should not be over-interpreted.

{This workflow works best when the dataset contains enough observations to constrain the model, the displayed uncertainties provide reasonable approximate weights, and the observation period covers a meaningful part of the decay. Small datasets can produce irregular profile intervals, weak residual tests, and unstable bootstrap intervals. For low count rates, researchers should use a Poisson model when the original counts and measurement times are available. A fit to plotted count rates then provides only an approximation. The source study~\cite{spillane2007} expects a single exponential decay, so we use a single-exponential model here. We have not tested the workflow on multiexponential decay because that analysis would require additional model components and identifiability checks and falls outside the scope of this study.}
	
{Regression and profile likelihood methods can use the actual time coordinates when the measurements have unequal spacing. However, the classical Fourier transform method and the simplest mean-displaced-ratio method assume equally spaced data. For the slightly uneven digitized time grid, we interpolated the data onto a uniform grid before applying the Fourier transform and used the mean reconstructed time difference for each displacement in the mean-displaced-ratio analysis. We therefore present the workflow as a set of related diagnostic tools, not as a fixed method for every reduced dataset.}

In future work, the same hierarchy of tests should be applied to long-duration, high-statistics measurements where the physical decay trend is shallow compared with normalization and detector stability effects. A natural example is the recent DEAP-3600 measurement of the \Ar{} half-life. The collaboration reports \(T_{1/2}(^{39}\mathrm{Ar})=(302\pm8_{\mathrm{stat}}\pm6_{\mathrm{syst}})~\mathrm{y}\)~\cite{adhikari2025}, while an earlier re-analysis of historical measurements gave \(T_{1/2}=268.2^{+3.1}_{-2.9}~\mathrm{y}\)~\cite{Golovko2023Ar39MFV}. This difference does not identify the origin of the tension, but it makes \Ar{} a useful target for transparent toy Monte Carlo studies and profile likelihood analyses of digitized or reduced data products.

Such studies would not replace collaboration-level detector analyses. They would ask a narrower question: which parts of the half-life inference are already testable from published or reduced information, and which parts require the full detector model. Synthetic datasets could be generated with known half-life, background, efficiency, pile-up, and slow-drift terms, and then analyzed using the same regression, profile likelihood, reparameterization, and ratio-based diagnostics used here. This type of reduced-data study is especially relevant when raw data are too large or too dependent on collaboration-specific software to be reused easily by outside analysts. The DEAP-3600 data volume is reported to be about \(150~\mathrm{TB}\)~\cite{adhikari2025}. Recent HPGe radiation sensor work demonstrated the pairwise--MFV approach on a long, stable decay dataset and identified short-span, high-statistics trigger-rate measurements, such as DEAP-3600, as natural targets for dedicated toy Monte Carlo and model-specific diagnostics~\cite{sensors2026pairwiseMFV}.	

		\section{
Conclusions
}
	\label{sec:conclusions}
	
This study demonstrates a reproducible method for testing half-life estimates when only plotted decay data are available. The analysis does not remeasure the \Au{} half-life from raw spectra and does not revise the recommended nuclear data. It shows what can be checked from a digitized figure and where the information’s limitations become important.	
	
For the selected \(T=293~\mathrm{K}\) \Au{} decay data, the weighted no-offset exponential fit gives \(T_{1/2}=(2.6687\pm0.0171)~\mathrm{d}\), which is in close agreement with the published individual curve value of \((2.669\pm0.017)~\mathrm{d}\). The profile likelihood interval, \(T_{1/2}=[2.6546,2.6830]~\mathrm{d}\), is consistent with the local covariance estimate. Therefore, the fit reproduces the published half-life scale, but the fitted normalization and half-life are strongly correlated. The digitized residuals are also broader than expected from the assigned point uncertainties alone, with \(\chi^2/\mathrm{ndf}=1.444\).	
	
The dominant figure-level sensitivity is a baseline-like count rate offset. A uniform offset of \(\pm0.05~\mathrm{cps}\) moves the fitted half-life by as much as \(0.0540~\mathrm{d}\). This shift is reported as a figure-level sensitivity scale and not as a calibrated systematic uncertainty of the original experiment. The signed residual offset profile gives the same message in a different way: over the finite time window of the digitized dataset, a constant residual offset, the fitted normalization, and the decay constant are partly degenerate. The offset profile minima are therefore diagnostic of non-identifiability and should not be interpreted as measured physical backgrounds or as revised half-life estimates.	
	
The pairwise ratio, MFV, MDR, toy Monte Carlo, and Fourier transform checks support the same cautious interpretation. They are useful because they test the reconstruction from different angles. However, they are not used as primary alternative estimates. The toy diagnostics show that lower pairwise or MFV central values can occur even for clean finite-window exponential data. The extended toy control shows that a longer observation window reduces the normalization--lifetime and offset--lifetime degeneracies. The Fourier transform result provides a residual baseline scale check, not an independent proof of a preferred half-life.	
	
Therefore, the main conclusion is simple. Well-resolved plotted data points can preserve sufficient information to reproduce the reported regression half-life and diagnose important sensitivity modes. At the same time, the figure-level data cannot recover the full experimental uncertainty model, the original background treatment, or the detector-level corrections. The safest practice for secondary analyses of published decay data is to report the primary regression result separately from statistical uncertainty, figure-level sensitivity scales, and diagnostic robustness checks.	

\vspace{6pt}	
		
\dataavailability{The supporting data and analysis files are available on OSF at \url{https://doi.org/10.17605/OSF.IO/BGZ6N} (accessed on 31 July 2026). The repository contains the digitized Figure~2 datasets from Ref.~\cite{spillane2007}, the analysis scripts, console logs, generated tables and figures, digitization quality assurance outputs, close-pair diagnostics, digitization overlay checks, and point-picking jitter tests used to reproduce the numerical results reported in this manuscript.}
 
\acknowledgments{The author would like to thank colleagues for their helpful discussions on half-life fitting practice and robustness diagnostics. The author is especially grateful to Maria Filimonova for her support and assistance. The author also thanks the management and staff at Canadian Nuclear Laboratories for fostering an enabling environment for this study, particularly Genevieve Hamilton and David Yuke. Microsoft 365 Copilot, based on a GPT-5 reasoning model, was used to assist with spelling, grammar, wording, and organization of the manuscript text. The author reviewed, edited, and approved the final version of the manuscript and remains responsible for its content.}

\abbreviations{Abbreviations}{
The following abbreviations are used in this manuscript:\\

\noindent 
\begin{tabular}{@{}ll}
	ATLAS & A Toroidal LHC ApparatuS \\
{CI} & confidence interval \\
{CMS} & Compact Muon Solenoid \\
{cps} & counts per second \\
{DEAP-3600} & Dark Matter Experiment Using Argon Pulse‑shape Discrimination \\
{FFT} & fast Fourier transform \\
{GPT} & generative pre‑trained transformer \\
{HEPData} & High Energy Physics Data repository \\
{HPB} & hybrid parametric bootstrapping \\
{HPGe} & high‑purity germanium \\
{LHC} & Large Hadron Collider \\
{LS} & least squares \\
{MDR} & mean displaced ratio \\
{MFV} & most frequent value \\
{ndf} & number of degrees of freedom \\
{OSF} & Open Science Framework \\
{PB} & petabyte \\
{PDG} & Particle Data Group \\
{RMS} & root mean square \\
{TB} & terabyte \\
\end{tabular}}
	
\abbreviations{Nomenclature}{%
	This list assigns one meaning to each symbol used in this manuscript.
}

\begingroup
\small

\setlength{\LTleft}{0pt}
\setlength{\LTright}{0pt}

\setlength{\tabcolsep}{5pt}
\renewcommand{\arraystretch}{1.12}

\begin{longtable}{@{}
		>{\raggedright\arraybackslash}p{0.12\textwidth}
		>{\raggedright\arraybackslash}p{0.76\textwidth}
		@{}}
	
	\textbf{Symbol} & \textbf{Definition} \\
	\endfirsthead
	
	\multicolumn{2}{@{}l}{%
		\small\itshape Nomenclature, continued from the previous page%
	} \\[3pt]
	
	\textbf{Symbol} & \textbf{Definition} \\
	\endhead
	
	\multicolumn{2}{r@{}}{%
		\small\itshape Continued on the next page%
	} \\
	\endfoot
	
	\endlastfoot
	
	$A(0)$
	&
	Published extrapolated initial count rate. Units: cps.
	\\
	
	$A(t)$
	&
	Model-predicted count rate at time \(t\). Units: cps.
	\\
	
	$\hat{A}(t_i)$
	&
	Fitted model value at time \(t_i\). Units: cps.
	\\
	
	$A_0$
	&
	Fitted initial count rate at \(t=0\). Units: cps.
	\\
	
	$A_{\mathrm{obs}}(t)$
	&
	Observed count-rate model when used conceptually. Units: cps.
	\\
	
	$A_i$
	&
	Digitized observed count-rate point at time \(t_i\). Units: cps.
	\\
	
	$b$
	&
	Uniform additive offset applied in the sensitivity scans. Units: cps.
	\\
	
	$b_{\log}$
	&
	Logarithmic background parameter used to enforce
	\(B_{\mathrm{bg}}\geq 0\). Dimensionless.
	\\
	
	$B_{\min}$
	&
	Numerical lower bound for the nonnegative background
	parameterization. Units: cps.
	\\
	
	$B_0$
	&
	Offset-prior center used in Equation~\eqref{eq:offsetprior}.
	Units: cps.
	\\
	
	$B_{\mathrm{bg}}$
	&
	Nonnegative physical background term used only for diagnostic
	profiling, Equation~\eqref{eq:bgmodel}. Units: cps.
	\\
	
	$B_{\mathrm{off}}$
	&
	Signed residual baseline offset for background-subtracted data,
	Equation~\eqref{eq:offmodel}. Units: cps.
	\\
	
	$B_{\mathrm{FFT}}$
	&
	Residual baseline estimate obtained from the Fourier-transform
	diagnostic. Units: cps.
	\\
	
	$E_\gamma$
	&
	Gamma-ray energy. Units: keV or MeV.
	\\
	
	$F(\omega)$
	&
	Finite-window Fourier transform of the count-rate curve at angular
	frequency \(\omega\). Units: cps\,s when defined as a continuous
	integral.
	\\
	
	$F(0)$
	&
	Zero-frequency Fourier component, equal to the time integral of the
	count-rate curve over the finite window. Units: cps\,s.
	\\
	
	$h$
	&
	Nonzero Fourier harmonic index used in the FFT diagnostic.
	Dimensionless.
	\\
	
	$n_{\mathrm{pts}}$
	&
	Number of digitized data points used in the fit. Dimensionless.
	\\
	
	$N_{\mathrm{FFT}}$
	&
	Number of uniformly spaced grid points used in the Fourier-transform
	diagnostic after interpolation. Dimensionless.
	\\
	
	$\mathcal{N}_0$
	&
	Effective normalization parameter in the reparameterized model.
	Units: counts.
	\\
	
	$p$
	&
	Number of fitted parameters. Dimensionless.
	\\
	
	$r_i$
	&
	Normalized residual,
	\([A_i-\hat{A}(t_i)]/\sigma_i\). Dimensionless.
	\\
	
	$s$
	&
	Uncertainty scale factor,
	\(\sqrt{\chi^2/\mathrm{ndf}}\). Dimensionless.
	\\
	
	$t$
	&
	Time since the start of counting. Units: s or d.
	\\
	
	$t_i$
	&
	Time coordinate of digitized point \(i\). Units: s or d.
	\\
	
	$t_{\min},t_{\max}$
	&
	Minimum and maximum reconstructed times. Units: s.
	\\
	
	$T$
	&
	Sample temperature. Units: K.
	\\
	
	$T_{\mathrm{win}}$
	&
	Finite observation window used in the Fourier-transform diagnostic.
	Units: s or d.
	\\
	
	$T_{1/2}$
	&
	Half-life. Units: s or d.
	\\
	
	$T_{1/2,\mathrm{FFT}}$
	&
	Half-life obtained from the first-harmonic FFT diagnostic.
	Units: d.
	\\
	
	$T_{1/2,\mathrm{reg}}$
	&
	Regression estimate of the half-life. Units: d.
	\\
	
	$T_{1/2,\mathrm{med}}$
	&
	Half-life obtained from the median of the pairwise lifetime
	estimates. Units: d.
	\\
	
	$T_{1/2,\mathrm{MFV}}$
	&
	Half-life obtained from the MFV summary of the pairwise lifetime
	estimates. Units: d.
	\\
	
	$T_{1/2,\mathrm{MDR}}$
	&
	Half-life obtained from the MDR smoothing diagnostic. Units: d.
	\\
	
	$w_i$
	&
	Least-squares weight, \(1/\sigma_i^2\).
	Units: cps\(^{-2}\).
	\\
	
	$\Delta t$
	&
	Nominal spacing between neighbouring time points. Units: s.
	\\
	
	$\Delta T_{1/2}$
	&
	Change in the fitted half-life under a perturbation.
	Units: s or d.
	\\
	
	$\Delta\chi^2$
	&
	Profile-likelihood difference,
	\(\chi^2-\chi^2_{\min}\). Dimensionless.
	\\
	
	$\lambda$
	&
	Exponential decay constant,
	\(\lambda=\ln 2/T_{1/2}=1/\tau\).
	Units: s\(^{-1}\) or d\(^{-1}\).
	\\
	
	$\rho$
	&
	Parameter correlation coefficient. Dimensionless.
	\\
	
	$\sigma_i$
	&
	Digitized point uncertainty. Units: cps.
	\\
	
	$\sigma'_i$
	&
	Rescaled point uncertainty, \(s\sigma_i\). Units: cps.
	\\
	
	$\sigma_B$
	&
	Offset-prior scale used in Equation~\eqref{eq:offsetprior}.
	Units: cps.
	\\
	
	$\sigma_{B,\mathrm{emp}}$
	&
	Empirical offset scale estimated near the minimum profile.
	Units: cps.
	\\
	
	$\sigma_{\mathrm{point}}$
	&
	Median point uncertainty used in residual-offset diagnostics.
	Units: cps.
	\\
	
	$\sigma_{\mathrm{res,rms}}$
	&
	Weighted residual RMS from the no-offset fit. Units: cps.
	\\
	
	$\chi^2$
	&
	Weighted sum of squared residuals. Dimensionless.
	\\
	
	$\chi^2_{\min}$
	&
	Minimum value of \(\chi^2\). Dimensionless.
	\\
	
	$\chi^2_\nu$
	&
	Reduced chi-square value reported in the source paper.
	Dimensionless.
	\\
	
	$\chi^2_{\mathrm{data}}$
	&
	Data contribution to the offset-profile objective function.
	Dimensionless.
	\\
	
	$\chi^2_{\mathrm{tot}}$
	&
	Total penalized objective function used in the weakly regularized
	offset fit. Dimensionless.
	\\
	
	$\omega$
	&
	Angular frequency used in Fourier-transform diagnostics.
	Units: rad\,s\(^{-1}\) or rad\,d\(^{-1}\).
	\\
	
	$\omega_h$
	&
	Nonzero Fourier angular frequency for harmonic \(h\),
	\(\omega_h=2\pi h/T_{\mathrm{win}}\).
	Units: rad\,s\(^{-1}\) or rad\,d\(^{-1}\).
	\\
	
	$A_{0,\mathrm{toy}}$
	&
	Fitted initial count rate for the 78-point toy dataset.
	Units: cps.
	\\
	
	$D_{\mathrm{KS}}$
	&
	Kolmogorov--Smirnov distance used in residual-distribution
	comparisons. Dimensionless.
	\\
	
	$p_{\mathrm{SW}},p_{\mathrm{AD}},p_{\mathrm{LF}}$
	&
	Shapiro--Wilk, Anderson--Darling, and Lilliefors
	normality-test \(p\)-values. Dimensionless.
	\\
	
	$p_{\mathrm{KS}},p_{\mathrm{WMW}}$
	&
	Kolmogorov--Smirnov and Wilcoxon--Mann--Whitney test
	\(p\)-values. Dimensionless.
	\\
	
	$\hat{B}_{\mathrm{bg}}(T_{1/2})$
	&
	Profiled nonnegative-background solution at a fixed half-life.
	Units: cps.
	\\
	
	$\hat{B}_{\mathrm{off}}(T_{1/2})$
	&
	Profiled signed-offset solution at a fixed half-life.
	Units: cps.
	\\
	
	$\tau$
	&
	Mean lifetime,
	\(\tau=T_{1/2}/\ln 2=1/\lambda\).
	Units: s or d.
	\\
	
	$\tau_{ij}$
	&
	Pairwise lifetime estimate from count-rate points \(i\) and \(j\).
	Units: d.
	\\
	
	$\tau_{\mathrm{reg}}$
	&
	Regression estimate of the mean lifetime. Units: d.
	\\
	
	$\tau_{\mathrm{med}}$
	&
	Median of the pairwise lifetime estimates. Units: d.
	\\
	
	$\tau_{\mathrm{MFV}}$
	&
	MFV summary of the pairwise lifetime estimates. Units: d.
	\\
	
	$q$
	&
	Time displacement, or lag, used in the MDR ratio
	\(A(t+q)/A(t)\). Units: s or d.
	\\
	
	$\ell$
	&
	Integer displacement index used in the discrete MDR calculation.
	Dimensionless.
	\\
	
	$L$
	&
	Maximum MDR displacement index controlling the smoothing length.
	Dimensionless.
	\\
	
	$M$
	&
	Highest observation index in the Dyson--Isenberg MDR notation,
	where the data contain \(M+1\) points. Dimensionless.
	\\
	
	$Y_\ell$
	&
	Mean displaced-ratio value at displacement index \(\ell\).
	Dimensionless.
	\\
	
	\(\mathrm{ndf}\)
	&
	Degrees of freedom,
	\(n_{\mathrm{pts}}-p\). Dimensionless.
	\\
	
\end{longtable}

\endgroup

	\appendixtitles{yes}
	\appendixstart
	\appendix
\renewcommand{\thetable}{A}
\renewcommand{\thefigure}{A}
	\section{
Mean Displaced Ratio Diagnostic
}
	\label{app:mdr}
	
As an additional ratio-based robustness check, we applied mean displaced ratio (MDR) smoothing~\cite{DysonIsenberg1971MDR} to the digitized \Au{} decay data. MDR smoothing was originally introduced as a way to reduce high-frequency noise in transient exponential data while preserving the underlying exponential time constants under ideal assumptions~\cite{IkossiAnastasiouRoenker1987}. In this study, MDR is used only as a diagnostic and not as the preferred half-life estimator.
	
The simplest idea for a clean single-exponential count rate curve is
	\begin{equation}
		A(t)=A_0\exp(-t/\tau),
		\label{eq:mdr_single_exp}
	\end{equation}
where \(A_0\) is the initial count rate scale and \(\tau\) is the mean lifetime. If the same curve is evaluated twice, separated by a lag \(q\), the ratio is
	\begin{equation}
		\frac{A(t+q)}{A(t)}
		=
		\frac{A_0\exp\left[-(t+q)/\tau\right]}
		{A_0\exp(-t/\tau)}
		=
		\exp(-q/\tau).
		\label{eq:mdr_single_exp_ratio}
	\end{equation}
{This} equation shows the main reason for using MDR as a diagnostic: the unknown normalization \(A_0\) cancels out exactly. In the reparameterized form \(A(t)=(\mathcal{N}_0/\tau)\exp(-t/\tau)\), the factor \(\mathcal{N}_0/\tau\) also cancels from the ratio. Therefore, the transformed quantity depends only on the time lag \(q\) and the decay lifetime. This removes the explicit fitted normalization from the MDR data and provides a useful check on the normalization--lifetime coupling observed in the regression fits. For equally spaced data, \(q=\ell\Delta t\), where \(\ell\) is an integer displacement index.
	
The distinct MDR calculation extends the ratio idea to many pairs of points. 
Following the Dyson--Isenberg indexing convention~\cite{DysonIsenberg1971MDR}, let the data contain \(M+1\) observations indexed by \(j=0,\ldots,M\). 
For each displacement index \(\ell\), the MDR value was calculated by averaging all available count rate ratios separated by that displacement:
\begin{equation}
	Y_\ell =
	\frac{1}{M-L+1}
	\sum_{j=0}^{M-L}
	\frac{A_{j+\ell}}{A_j},
	\qquad
	\ell=0,1,\ldots,L .
	\label{eq:mdr_discrete}
\end{equation}
{Here} \(A_j\) is the digitized count rate at point \(j\), and \(L\) sets the maximum displacement used in the MDR curve. 
With this definition, \(Y_0=1\). 
Smaller \(L\) values include more ratios in each average and give stronger smoothing, but they shorten the displacement range. 
Larger \(L\) values keep a longer displacement range but average fewer ratios.

For an ideal single exponential, Equation~\eqref{eq:mdr_single_exp_ratio} predicts that the averaged MDR curve should also follow an exponential form. 
The MDR values were therefore fitted with
\begin{equation}
	Y(q)=\exp(-q/\tau)
	=
	\exp\left(-\ln 2\,\frac{q}{T_{1/2}}\right),
	\label{eq:mdr_fit}
\end{equation}
where \(q\) is the time displacement associated with \(\ell\). 
We did not include a free amplitude in the fit because the MDR curve is normalized by construction. The digitized time spacing is close to, but not exactly, uniform, so the mean reconstructed time displacement for each \(\ell\) was used in the fit.
	
The source \Au{} dataset analyzed in this work was already background-subtracted in the original publication. Therefore, MDR was applied to net count rate data rather than to raw count rates with an unknown physical background. This distinction is important because a constant background would not cancel the simple ratio \(A(t+q)/A(t)\). In this application, MDR is interpreted only as a smoothing and normalization-removal diagnostic. The signed residual offset profile remains a more direct test of baseline sensitivity.
	
The MDR was evaluated for four smoothing lengths,
	\[
L/M = 0.247,\quad 0.494,\quad 0.753,\quad 0.896 .
	\]
{The} fitted half-lives were \(2.7012,\ 2.6572,\ 2.6766,\) and \(2.7049~\mathrm{d}\), corresponding to shifts of \(+1.217\%\), \(-0.432\%\), \(+0.296\%\), and \(+1.356\%\) relative to the no-offset regression value \(T_{1/2}=2.6687~\mathrm{d}\). The closest agreement was obtained for \(L/M=0.753\), with
	\[
	T_{1/2,\mathrm{MDR}}=2.6766~\mathrm{d},
	\]
a difference in \(0.0079~\mathrm{d}\), or \(0.296\%\).
	
The MDR fits should not be interpreted as independent high-precision estimates. The MDR points share the same original digitized observations and are therefore correlated. This correlation affects the obvious goodness-of-fit and the formal fit uncertainties. For example, the MDR fits gave \(\chi^2/\mathrm{ndf}\) values ranging from \(0.051\) to \(1.586\), reflecting the transformed curve’s diagnostic nature rather than independent residual statistics. Nevertheless, the MDR half-lives remained close to the no-offset regression scale. Therefore, the MDR smoothing check supports the conclusion that the reconstructed half-life scale is stable under a ratio-based smoothing transformation, while the profile-based residual offset analysis remains the preferred method for assessing baseline sensitivity.
	
	As an implementation and sampling-regularity control, the same MDR procedure was also applied to the 78-point and 234-point toy datasets. These toy datasets are closer to the distinct MDR formula assumptions because they are sampled on idealized regular time grids. For the 78-point toy dataset, the MDR half-lives ranged from \(2.6518\) to \(2.7321~\mathrm{d}\), corresponding to shifts of \(-0.624\%\) to \(+2.385\%\) relative to that dataset's no-offset regression value. For the extended 234-point toy dataset, the MDR half-lives ranged from \(2.6636\) to \(2.6973~\mathrm{d}\), corresponding to shifts of \(-0.149\%\) to \(+1.112\%\). Therefore, the extended toy control shows improved MDR stability on a cleaner and more densely sampled grid. These toy-data MDR results are used only as implementation checks and are not treated as additional experimental constraints on the \Au{} half-life.
	
	\section{Auxiliary Model Checks}
	\label{app:auxiliary_checks}
This appendix collects two checks that support the interpretation of the main analysis but are not used to define the primary half-life estimate. The first check changes the parameterization of the no-offset exponential model. The second check adds a constrained nonnegative background term. Both tests are useful for understanding parameter coupling, but neither replaces the weighted regression and profile likelihood results.

The equivalent normalization check rewrites the model
	\[
	A(t)=A_0\exp(-t/\tau)
	\]
as
	\[
	A(t)=\frac{\mathcal{N}_0}{\tau}\exp(-t/\tau),
	\]
where \(\tau=T_{1/2}/\ln2\) is the mean lifetime and \(\mathcal{N}_0=A_0\tau\) in the chosen time unit. This parameterization gives the same half-life and equivalent initial rate, \(A_0=3.7333~\mathrm{cps}\). The fitted normalization scale is \(\mathcal{N}_0=1.2419\times10^6\) when \(\tau\) is expressed in seconds. This scale is only a count rate normalization and should not be interpreted as the number of radioactive atoms or source decays unless the detector efficiency, gamma-ray emission probability, counting geometry, live time, and related corrections are explicitly included~\cite{Golovko2022Efficiency,Golovko2024Scintillation}.

The reparameterized fit confirms that the scale--lifetime coupling is not removed by changing notation. The correlation between \(\mathcal{N}_0\) and \(T_{1/2}\) is \(0.936\), whereas the original \(A_0\)--\(T_{1/2}\) correlation is \(-0.831\). The sign changes because the normalization is defined differently, but the main message remains unchanged: over the finite digitized time window, the scale parameter and decay constant are strongly linked.

For completeness, the following nonnegative background model was also considered:
	\begin{equation}
		A(t)=A_0\exp\left(-\ln 2\,\frac{t}{T_{1/2}}\right)+B_{\mathrm{bg}}, \qquad B_{\mathrm{bg}}\ge 0 .
		\label{eq:bgmodel}
	\end{equation}
{This} model is physically natural for raw count rate data. However, the curve studied here was already background-subtracted. Therefore, it is only used as a conservative stress test. We enforced non-negativity in the nonlinear fits by writing \(B_{\mathrm{bg}}=\exp(b_{\log})\) with a finite lower bound for numerical stability.

The nonnegative-background profile has its minimum at the no-offset half-life, \mbox{\(T_{1/2}=2.6687~\mathrm{d}\)}. Therefore, allowing a nonnegative additive term does not produce a meaningful shift in the fitted half-life. The useful information from this check is the boundary behavior: for part of the profile, the fitted background is driven to zero. In that boundary-pinned region, the likelihood is non-regular and standard \(\Delta\chi^2\) interval rules should not be used. Therefore, the nonnegative-background model is treated only as an auxiliary diagnostic, whereas the signed residual offset profile is used as the main baseline-sensitivity test for the processed digitized dataset.

\section{Fourier Transform Baseline Offset Diagnostic}
\label{app:fft_baseline}

As an additional residual-baseline diagnostic, we tested the Fourier transform method for a single exponential plus constant offset. This method was introduced for the analysis of single-exponential transients in deep-level transient spectroscopy, where the ratio of the real and imaginary parts of a nonzero Fourier component gives the exponential decay constant independent of the amplitude and constant baseline offset~\cite{Kirchner1981FFT}. The same idea was later used as a baseline-extraction step in refined method-of-moment analyses of multiexponential transients~\cite{IkossiAnastasiouRoenker1987}.

For this diagnostic, the count rate curve was written as follows:
\[
A(t)=A_0 e^{-\lambda t}+B,
\]
where \(A(t)\) is the count rate at time \(t\), \(A_0\) is the initial count rate scale, \(\lambda\) is the decay constant, and \(B\) is a constant residual offset. Over a finite observation window \(0\le t<T_{\mathrm{win}}\), the Fourier transform is defined using the convention
\[
F(\omega)=\int_0^{T_{\mathrm{win}}} A(t)e^{-i\omega t}\,dt .
\]
{At} Fourier frequencies
\[
\omega_h=\frac{2\pi h}{T_{\mathrm{win}}},
\qquad h=1,2,\ldots,
\]
the constant term does not contribute because
\[
\int_0^{T_{\mathrm{win}}} B e^{-i\omega_h t}\,dt=0 .
\]
{Thus}, for nonzero harmonics,
\[
F(\omega_h)
=
A_0\int_0^{T_{\mathrm{win}}} e^{-(\lambda+i\omega_h)t}\,dt
=
A_0
\frac{1-e^{-\lambda T_{\mathrm{win}}}}
{\lambda+i\omega_h}.
\]
{Writing}
\[
C=A_0\left(1-e^{-\lambda T_{\mathrm{win}}}\right),
\]
we obtain
\[
F(\omega_h)
=
\frac{C}{\lambda+i\omega_h}
=
\frac{C(\lambda-i\omega_h)}
{\lambda^2+\omega_h^2}.
\]
{Therefore},
\[
\Re[F(\omega_h)]
=
\frac{C\lambda}{\lambda^2+\omega_h^2},
\qquad
\Im[F(\omega_h)]
=
-\frac{C\omega_h}{\lambda^2+\omega_h^2}.
\]
{The} ratio gives
\[
\lambda
=
-\omega_h
\frac{\Re[F(\omega_h)]}{\Im[F(\omega_h)]}.
\]
{Once} \(\lambda\) is known, the amplitude follows the next equation:
\[
A_0
=
\Re[F(\omega_h)]
\frac{\lambda^2+\omega_h^2}
{\lambda\left(1-e^{-\lambda T_{\mathrm{win}}}\right)}.
\]
{The} residual offset is obtained from the zero-frequency component as follows:
\[
F(0)
=
\int_0^{T_{\mathrm{win}}} A(t)\,dt
=
A_0
\frac{1-e^{-\lambda T_{\mathrm{win}}}}{\lambda}
+
B T_{\mathrm{win}} .
\]
{Solving} for \(B\) gives
\[
B_{\mathrm{FFT}}
=
\frac{F(0)}{T_{\mathrm{win}}}
-
A_0
\frac{1-e^{-\lambda T_{\mathrm{win}}}}
{\lambda T_{\mathrm{win}}}.
\]
{For} an unscaled discrete FFT, we replace \(F(0)/T_{\mathrm{win}}\) with the mean of the sampled signal, \(\mathrm{FFT}(0)/N_{\mathrm{FFT}}\).

{The standard Fourier transform calculation needs equally spaced data. We first sorted the digitized points by time and set the first reconstructed time to zero. We then created 1024 equally spaced time points over the same observation period. We left out the final endpoint so that the Fourier period matched the observation period. We used linear interpolation between neighboring digitized points to estimate the count rate at each new time point. This step did not smooth the data or fit a new decay curve. It only created the equal spacing needed for the Fourier transform calculation. This approach follows the fixed-sampling requirement for mono-exponential fitting methods~\cite{Fuhrmann2014MonoExponential}. Interpolation may create correlations and affect the higher Fourier harmonics. The source article had also already subtracted the background from the plotted data. For these reasons, we use $B_{\mathrm{FFT}}$ only as a qualitative check of the residual baseline, not as a measured physical background~rate.}

For the digitized dataset, the first nonzero Fourier harmonic
\[
T_{1/2,\mathrm{FFT}}=2.67554~\mathrm{d},
\]
compared with the no-offset regression value
\[
T_{1/2}=2.66869~\mathrm{d}.
\]
{The} difference is
\[
\Delta T_{1/2}=0.00685~\mathrm{d},
\qquad
\frac{\Delta T_{1/2}}{T_{1/2}}=0.257\%.
\]
{The} corresponding residual baseline estimate was as follows:
\[
B_{\mathrm{FFT}}=-0.0319~\mathrm{cps}.
\]
{This} value is comparable to the weighted residual RMS of the no-offset fit, as follows:
\[
\sigma_{\mathrm{res,rms}}=0.0330~\mathrm{cps}.
\]

{Thus},  
the Fourier transform diagnostic estimates a residual baseline scale of 0.97~residual RMS units. This is much smaller than the offsets associated with the minima of the signed residual offset profiles, which were approximately
\[
B_{\mathrm{off}}\simeq -0.167~\mathrm{cps}
\]
for the unconstrained offset profile and
\[
B_{\mathrm{off}}\simeq -0.137~\mathrm{cps}
\]
for the weakly regularized offset profile. These larger profile offsets are accompanied by substantial shifts in the fitted half-life and are therefore better interpreted as part of the amplitude--lifetime--offset degeneracy than as direct estimates of a physical residual background.

\textls[-15]{The first Fourier harmonic gives a half-life close to the no-offset regression result, indicating  a dominant single-exponential trend over the digitized time window. However, higher Fourier harmonics were unstable or nonphysical, indicating sensitivity to finite-window, interpolation, and digitization effects. Therefore, the FFT result is retained only as a qualitative residual baseline check. The primary evidence for the half-life estimate comes from the weighted regression, residual analysis, profile likelihood, and ratio-based~diagnostics.}
	
\section{Extended Toy Monte Carlo Diagnostic}
\label{app:extended_toy_control}
\setcounter{table}{0}
\setcounter{figure}{0}

\renewcommand{\thetable}{D\arabic{table}}
\renewcommand{\thefigure}{D\arabic{figure}}

To separate finite-window effects from estimator-specific behavior, the toy control was repeated with the same one-hour sampling but three times the observation duration. The extended toy dataset contains 234 points and covers approximately \(9.7~\mathrm{d}\), compared with about \(3.2~\mathrm{d}\) for the 78-point toy dataset. This keeps the decay model fixed while giving the exponential curve a longer time baseline over which \(A_0\), \(T_{1/2}\), and possible baseline-like offsets can be separated.

The no-offset regression remained well-behaved for the extended toy data.
\[
T_{1/2,\mathrm{reg,ext}}=(2.6676\pm0.0121)~\mathrm{d},
\qquad
\chi^2/\mathrm{ndf}=0.977,
\qquad
s=0.988 .
\]
{The}  
longer time window reduced the (\(A_0\),\(T_{1/2}\)) correlation from about \(-0.83\) in the \mbox{78-point} toy case to \(-0.75\). The signed-offset profile also changed in the expected direction: the free and weak-prior offset fits stayed close to the no-offset solution, near \mbox{\(T_{1/2}\simeq2.67\)--\(2.68~\mathrm{d}\)}, with fitted offsets of only about \(-0.004\) -- \(-0.005~\mathrm{cps}\). This indicates that the broad offset--lifetime valley seen in the shorter datasets is mainly a finite-window identifiability effect. With a longer observation window, a constant offset and a change in half-life are more clearly separated.

Table~\ref{tab:toy-extended-control-summary} shows how a longer baseline time changes the diagnostics. The regression uncertainty is reduced from \(\pm2241~\mathrm{s}\) in the 78-point toy data to \(\pm1046~\mathrm{s}\) in the extended toy data, and the leave-one-out and early/late truncation sensitivities also decrease. The signed-offset behavior improves in the same direction: with a longer observation window, the offset--lifetime trade-off is weaker because a constant baseline term and a change in half-life are more clearly separated.

\begin{table}[H]
\renewcommand{\thetable}{A\arabic{table}}
	\centering
	\caption{{Extended}  toy Monte Carlo control diagnostics for the no-offset fit. The extended toy dataset contains 234 points on an idealized uniform \(3600~\mathrm{s}\) time grid, corresponding to three times the duration of the 78-point toy dataset. The entries are diagnostic shifts relative to the extended-toy no-offset fit, not systematic uncertainties.}
	\label{tab:toy-extended-control-summary}
	\small
\centering 

		\begin{tabularx}{\textwidth} {lLLL}
		\toprule
		\textbf{Component or Diagnostic} & \textbf{\(\bm{T_{1/2}}\) Shift [s]} & \textbf{\(\bm{T_{1/2}}\) Shift [d]} &\textbf{ Relative [\%]} \\
		\midrule
		Central value from extended-toy no-offset fit & 230{,}483.1 & \(2.6676\) & -- \\
		Covariance matrix standard error & \(\pm1046\) & \(\pm0.0121\) & \(\pm0.454\) \\
		Profile interval, \(\Delta\chi^2=1\) & \(-1055,\,+1061\) & \(-0.0122,\,+0.0123\) & \(-0.458,\,+0.460\) \\
		Absolute time shift, up to \(\pm60~\mathrm{s}\) & \(\le1.8\times10^{-5}\) & \(\le2.1\times10^{-10}\) & negligible \\
		Time-scale increase, \(+0.5\%\) & \(+1152\) & \(+0.0133\) & \(+0.500\) \\
		Uniform offset, \(-0.05 \le b \le 0.05~\mathrm{cps}\) & \(8607.8\) & \(0.0996\) & \(3.73\) \\
		Constant uncertainty check, \(\sigma=\mathrm{median}(\sigma_i)\) & \(342.7\) & \(0.00397\) & \(0.149\) \\
		Leave-one-out maximum & \(355\) & \(0.00411\) & \(0.154\) \\
		Drop earliest \(10\%\) of points & \(-197\) & \(-0.00228\) & \(-0.0855\) \\
		Drop latest \(10\%\) of points & \(-17.4\) & \(-0.000201\) & \(-0.00755\) \\
		\bottomrule
	\end{tabularx}
\end{table}

The only diagnostic increase is the fixed uniform count rate offset scan. This increase should be carefully interpreted. The imposed offset range, \(\pm0.05~\mathrm{cps}\), is the same in absolute size as in the shorter toy test, but it becomes a stronger perturbation at late times because the extended decay data reaches lower count rates. Therefore, this row should be read as a stress test of baseline sensitivity, not as a realistic systematic uncertainty.

Overall, the extended toy test supports the interpretation that much of the regression and offset behavior, and part of the pairwise/MFV central-value shift, comes from the finite observation window. When the time baseline is extended, the regression becomes better constrained, the signed-offset degeneracy weakens, and the pairwise median and MFV estimates approach the regression result. Therefore, the lower pairwise central values observed in the shorter datasets should not be interpreted by themselves as evidence of a different physical half-life.

The extended toy dataset was also processed using the pairwise-ratio and MFV workflow. The pairwise median gives
\[
\tau_{\mathrm{med,ext}}=3.8282^{+0.0323}_{-0.0289}~\mathrm{d},
\qquad
T_{1/2,\mathrm{med,ext}}=2.6535^{+0.0224}_{-0.0200}~\mathrm{d},
\]
while the pairwise MFV gives
\[
\tau_{\mathrm{MFV,ext}}=3.8289^{+0.0263}_{-0.0259}~\mathrm{d},
\qquad
T_{1/2,\mathrm{MFV,ext}}=2.6540^{+0.0183}_{-0.0180}~\mathrm{d}.
\]
{Both} the central values remain slightly lower than those of the extended-toy regression result,
\[
\tau_{\mathrm{reg,ext}}=3.8486^{+0.0175}_{-0.0175}~\mathrm{d},
\qquad
T_{1/2,\mathrm{reg,ext}}=2.6676^{+0.0121}_{-0.0121}~\mathrm{d},
\]
but the 68.27\% intervals overlap.
Compared with the 78-point toy dataset, the pairwise-to-regression shifts are smaller:
\[
\tau_{\mathrm{med,ext}}-\tau_{\mathrm{reg,ext}}=-0.0204~\mathrm{d},
\qquad
\tau_{\mathrm{MFV,ext}}-\tau_{\mathrm{reg,ext}}=-0.0197~\mathrm{d}.
\]

{This}  
reduction shows that the downward pairwise/MFV shift is partly a finite-window effect. The exponential shape is better constrained with a longer observation window, the offset--lifetime degeneracy is reduced, and the pairwise estimators move closer to the regression result.

\begin{figure}[H]
\renewcommand{\thefigure}{A\arabic{figure}}
	\includegraphics[width=0.99\linewidth]{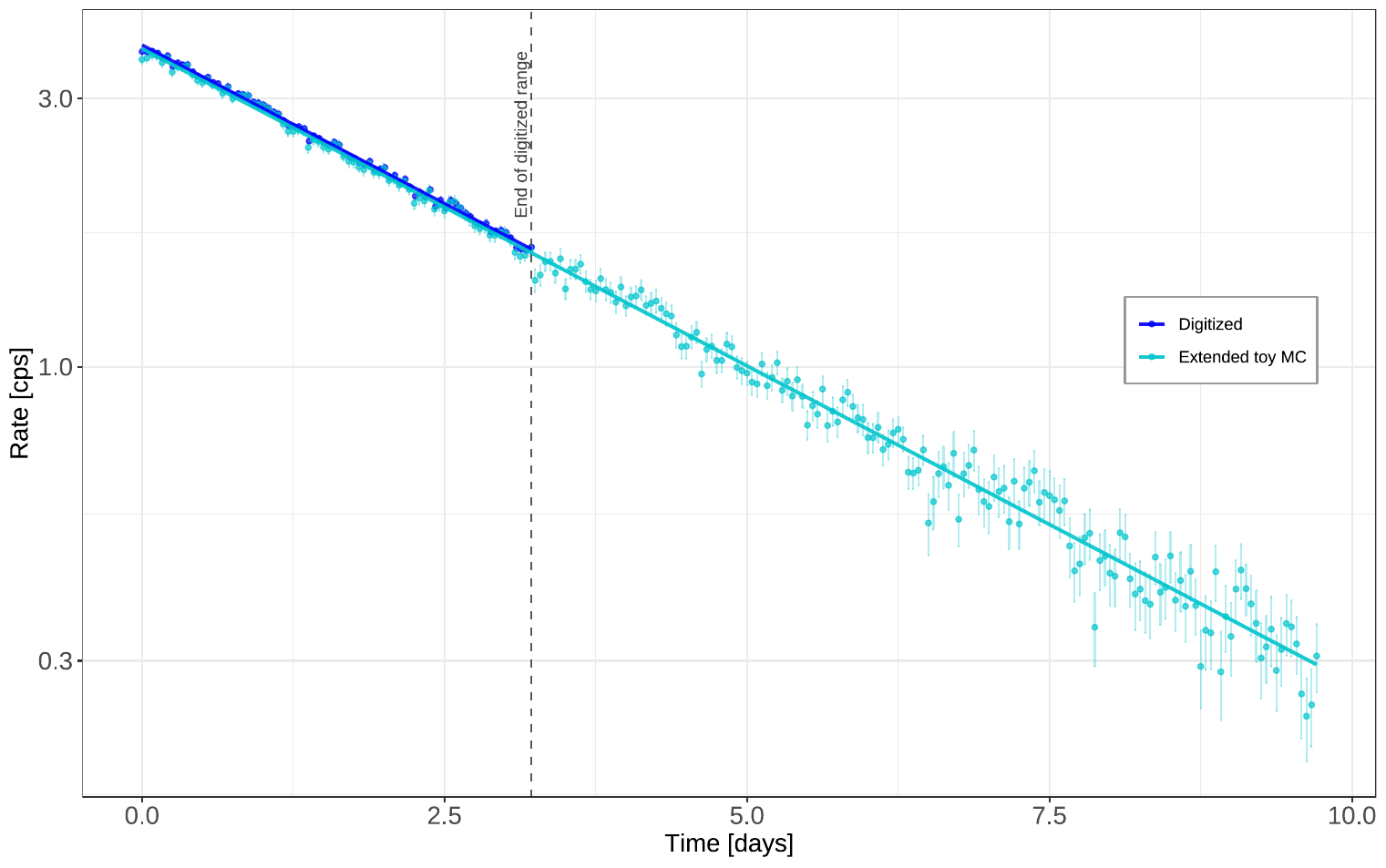}
	\caption{Comparison of the digitized \Au{} data points and the 3-times extended toy dataset under the no-offset exponential regression model. The digitized data cover about \(3.2~\mathrm{d}\), whereas the extended toy dataset continues to about \(9.7~\mathrm{d}\) on a uniform \(3600~\mathrm{s}\) time grid. The agreement over the common time range shows that the toy control is centered on the same decay scale, while the longer time baseline allows the regression, residual offset, and pairwise/MFV diagnostics to be tested under better-constrained conditions.}
	\label{fig:digitized-extended-toy-fit-overlay}
\end{figure}

Figure~\ref{fig:digitized-extended-toy-fit-overlay} compares the digitized \Au{} curve with the 3-times extended toy dataset under the same no-offset exponential regression model. Over the common time range, the two fitted curves follow the same decay scale. The longer toy window then provides a controlled way to examine effects that cannot be separated using the digitized figure alone, such as the normalization--lifetime correlation, the signed-offset degeneracy, and the finite-window behavior of the pairwise median and MFV estimators. This toy comparison is an idealized control that shows how the analysis behaves when the same decay model is observed over a longer baseline time.

As a descriptive residual diagnostic, the normalized residuals from the digitized and extended-toy fits were compared after subtracting the fitted exponential trend. The Kolmogorov--Smirnov diagnostic compares the overall shapes of the two residual distributions, and the Wilcoxon--Mann--Whitney diagnostic checks whether one residual distribution is shifted relative to the other. The Kolmogorov--Smirnov result is \(D_{\mathrm{KS}}=0.1026\) with \(p_{\mathrm{KS}}=0.570\), and the Wilcoxon--Mann--Whitney result is \(p_{\mathrm{WMW}}=0.959\). These \(p\)-values are much larger than the commonly used reference level \(\alpha=0.05\), which corresponds to a 5\% false-alarm threshold for calling a difference statistically clear. Therefore, these descriptive tests do not show a clear difference in the residual shape or central location. The median normalized residuals are close to zero for both datasets, \(0.0379\) for the digitized data and \(-0.0353\) for the extended toy data. Therefore, the main difference is not where the residual distributions are centered but how wide they are: the digitized residuals are broader, consistent with the larger goodness-of-fit scale factor in the digitized reconstruction.

		\reftitle{References}

\end{document}